\documentclass[a4paper,12pt]{article}
\usepackage{graphicx}
\begin{document}
%%%%%%%%%Fawzi's Short CUT: GENERAL%%%%%%%%%%%%%%%%%
%\bibliographystyle{unsrt}
\bibliographystyle{lesHouches}

\newcommand{\beqn}{\begin{eqnarray}}
\newcommand{\eeqn}{\end{eqnarray}}
\newcommand{\ra}{\rightarrow}

\newcommand{\np}{Nucl.\,Phys.\,}
\newcommand{\pl}{Phys.\,Lett.\,}
\newcommand{\pr}{Phys.\,Rev.\,}
\newcommand{\prl}{Phys.\,Rev.\,Lett.\,}
\newcommand{\prep}{Phys.\,Rep.\,}
\newcommand{\nuclinst}{{\em Nucl.\ Instrum.\ Meth.\ }}
\newcommand{\annp}{{\em Ann.\ Phys.\ }}
\newcommand{\intjmp}{{\em Int.\ J.\ of Mod.\  Phys.\ }}

%GENERAL

\newcommand{\mw}{M_{W}}
\newcommand{\mww}{M_{W}^{2}}
\newcommand{\mwmw}{M_{W}^{2}}

\newcommand{\mz}{M_{Z}}
\newcommand{\mzz}{M_{Z}^{2}}

\newcommand{\cw}{\cos\theta_W}
\newcommand{\sw}{\sin\theta_W}
\newcommand{\tw}{\tan\theta_W}
\def\cww{\cos^2\theta_W}
\def\sww{\sin^2\theta_W}
\def\tww{\tan^2\theta_W}

\def\noi{\noindent}
\def\nn{\noindent}

\def\sinb{\sin\beta}
\def\cosb{\cos\beta}
\def\sinbb{\sin (2\beta)}
\def\cosbb{\cos (2 \beta)}
\def\tgb{\tan \beta}
\def\tgbt{$\tan \beta\;\;$}
\def\tgbsq{\tan^2 \beta}
\def\sel{\tilde{e}_L}
\def\ser{\tilde{e}_R}
\def\msel{m_{\sel}}
\def\mser{m_{\ser}}
\def\mslr{m_{\tilde{l}_R}}
\def\m0{M_0}
\def\amu{\delta a_\mu}

\def\slashE{E\kern -.620em {/}}
%%%%%%%%%%%%%%%%%%%%%

\def\mchi{m_{\chi^+}}
\def\neuto{\tilde{\chi}_1^0}
\def\mneuto{m_{\tilde{\chi}_1^0}}
\def\neutt{\tilde{\chi}_2^0}
\def\neutth{\tilde{\chi}_3^0}
\def\ma{M_A}
\def\mstau{m_{\tilde\tau}}
\def\msne{m_{\tilde\nu}}
\def\msnee{m_{{\tilde\nu}_e}}
\def\mh{M_h}

\def\sinb{\sin\beta}
\def\cosb{\cos\beta}
\def\sinbb{\sin (2\beta)}
\def\cosbb{\cos (2 \beta)}
\def\tgb{\tan \beta}
\def\tgbt{$\tan \beta\;\;$}
\def\tgbsq{\tan^2 \beta}
\def\sinal{\sin\alpha}
\def\cosal{\cos\alpha}
%%%%%%%%%%%%%%%%%%%%%%%%5
\def\stop{\tilde{t}}
\def\sto{\tilde{t}_1}
\def\stt{\tilde{t}_2}
\def\stl{\tilde{t}_L}
\def\str{\tilde{t}_R}
\def\msto{m_{\sto}}
\def\mstosq{m_{\sto}^2}
\def\mstt{m_{\stt}}
\def\msttsq{m_{\stt}^2}
\def\mt{m_t}
\def\mtsq{m_t^2}
\def\sint{\sin\theta_{\stop}}
\def\sintt{\sin 2\theta_{\stop}}
\def\cost{\cos\theta_{\stop}}
\def\sintsq{\sin^2\theta_{\stop}}
\def\costsq{\cos^2\theta_{\stop}}
\def\mqtt{\M_{\tilde{Q}_3}^2}
\def\mutt{\M_{\tilde{U}_{3R}}^2}
%%%%%%%%%%%%%%%%%%%%%
\def\sbottom{\tilde{b}}
\def\sbo{\tilde{b}_1}
\def\sbt{\tilde{b}_2}
\def\sbl{\tilde{b}_L}
\def\sbr{\tilde{b}_R}
\def\msbo{m_{\sbo}}
\def\msbosq{m_{\sbo}^2}
\def\msbt{m_{\sbt}}
\def\msbtsq{m_{\sbt}^2}
\def\mt{m_t}
\def\mtsq{m_t^2}
%%%%%%%%%%%%%%%%%%%%%
\def\selectron{\tilde{e}}
\def\seo{\tilde{e}_1}
\def\set{\tilde{e}_2}
\def\sel{\tilde{e}_L}
\def\se1{\tilde{e}_1}
\def\ser{\tilde{e}_R}
\def\mseo{m_{\seo}}
\def\mseosq{m_{\seo}^2}
\def\mset{m_{\set}}
\def\msetsq{m_{\set}^2}
\def\msel{m_{\sel}}
\def\mser{m_{\ser}}
\def\mse1{m_{\se1}}
\def\me{m_e}
\def\mesq{m_e^2}
%%%%%%%%%%%%%%%%%%%%%
\def\snu{\tilde{\nu}}
\def\snue{\tilde{\nu_e}}
\def\set{\tilde{e}_2}
\def\snul{\tilde{\nu}_L}
\def\msnue{m_{\snue}}
\def\msnuesq{m_{\snue}^2}
%%%%%%%%%%%%%%%%%%%%%
\def\smuon{\tilde{\mu}}
\def\smul{\tilde{\mu}_L}
\def\smur{\tilde{\mu}_R}
\def\msmul{m_{\smul}}
\def\msmulsq{m_{\smul}^2}
\def\msmur{m_{\smur}}
\def\msmursq{m_{\smur}^2}
%%%%%%%%%%%%%%%%%%%%%%%%%%
\def\stau{\tilde{\tau}}
\def\stauo{\tilde{\tau}_1}
\def\staut{\tilde{\tau}_2}
\def\staul{\tilde{\tau}_L}
\def\staur{\tilde{\tau}_R}
\def\mstauo{m_{\stauo}}
\def\mstauosq{m_{\stauo}^2}
\def\mstaut{m_{\staut}}
\def\mstautsq{m_{\staut}^2}
\def\mtau{m_\tau}
\def\mtausq{m_\tau^2}
%%%%%%%%%%%%%%%%%%%%%%%%%%%%%%
\def\gluino{\tilde{g}}
\def\mgluino{m_{\tilde{g}}}
\def\mchi{m_\chi^+}
\def\neuto{\tilde{\chi}_1^0}
\def\mneuto{m_{\tilde{\chi}_1^0}}
\def\neutt{\tilde{\chi}_2^0}
\def\mneutt{m_{\tilde{\chi}_2^0}}
\def\neutth{\tilde{\chi}_3^0}
\def\mneutth{m_{\tilde{\chi}_3^0}}
\def\neutf{\tilde{\chi}_4^0}
\def\mneutf{m_{\tilde{\chi}_4^0}}
\def\chargop{\tilde{\chi}_1^+}
\def\chargopm{\tilde{\chi}_1^\pm}
\def\mchargo{m_{\tilde{\chi}_1^+}}
\def\chargtp{\tilde{\chi}_2^+}
\def\mchargt{m_{\tilde{\chi}_2^+}}
\def\chargom{\tilde{\chi}_1^-}
\def\chargtm{\tilde{\chi}_2^-}
\def\bino{\tilde{b}}
\def\wino{\tilde{w}}
\def\photino{\tilde{\gamma}}
\def\zino{tilde{z}}
%%%%%%%%%%%%%%%%%%%%%%%%%%%%%%%%%
\def\sdowno{\tilde{d}_1}
\def\sdownt{\tilde{d}_2}
\def\sdownl{\tilde{d}_L}
\def\sdownr{\tilde{d}_R}
\def\supo{\tilde{u}_1}
\def\supt{\tilde{u}_2}
\def\supl{\tilde{u}_L}
\def\supr{\tilde{u}_R}
%%%%%%%%%%Higgses masses%%%%%%%%%%%%
\def\mh{m_h}
\def\mht{m_h^2}
\def\MH{M_H}
\def\MHt{M_H^2}
\def\MA{M_A}
\def\MAt{M_A^2}
\def\MHp{M_H^+}
\def\MHm{M_H^-}
\def\mbmb{m_b(m_b)}
\def\epem{e^+e^-}
\def\epemt{$e^+e^-$}
\def\siginv{\sigma_{\gamma+inv}}
\def\gmuon{$(g-2)_\mu$}
\def\r12{r_{12}}
\def\bsgamma{b\ra s\gamma}
\def\bsmu{B_s\ra \mu^+\mu^-}
\def\feynhiggs{{\tt FeynHiggs}}
\def\micro{{\tt micrOMEGAs}}
\def\mhf{M_{1/2}}
\def\suspect{{\tt Suspect}}
\def\softsusy{{\tt SOFTSUSY}}
\def\xenon{{\tt Xenon}}
\def\zeplin{{\tt Zeplin}}
\def\zepliniv{{\tt ZeplinIV}}
\def\edelweiss{{\tt Edelweiss}}
\def\genius{{\tt Genius}}
\def\cdms{{\tt CDMS}}
\def\picasso{{\tt PICASSO}}
\def\simple{{\tt Simple}}
\def\naiad{{\tt Naiad}}
\def\dama{{\tt DAMA}}

\begin{titlepage}
\def\baselinestretch{1.2}
\vspace*{\fill}
\begin{center}
{\large {\bf {\em WMAP constraints on SUGRA  models with non-universal gaugino masses and prospects for direct detection.}}}
\vspace{1cm}

\begin{tabular}[t]{c}

{\bf G.~B\'elanger$^{1}$, F.~Boudjema$^{1}$, A.~Cottrant$^{1}$,  A. Pukhov$^{2}$, A. Semenov$^{3}$}
 \\
\\
\\
{\it 1. Laboratoire de Physique Th\'eorique}
{\large LAPTH}
\footnote{URA 14-36 du CNRS, associ\'ee  \`a
l'Universit\'e de Savoie.}\\
 {\it Chemin de Bellevue, B.P. 110, F-74941 Annecy-le-Vieux,
Cedex, France.}\\

{\it 2. Skobeltsyn Institute of Nuclear Physics,
Moscow State University} \\ {\it Moscow 119992,
Russia }\\

{\it 3. Joint Institute for Nuclear Research (JINR)}\\
{\it 141980, Dubna, Moscow Region, Russia}\\

\end{tabular}
\end{center}

\centerline{ {\bf Abstract} } \baselineskip=14pt \noindent
%%%%%%%%%%%%%%%%%%%%%%%%%%%%%%%%%%%%%%%%%%%%%%%%%%%%%%%%%%%%%%%%%%%%%
{\small We discuss constraints on supersymmetric models arising
from the relic density measurements of WMAP as well as from direct
and precision measurements, LEP, $\bsgamma$, $(g-2)_\mu$, $\bsmu$.
We consider mSUGRA models and their extensions  with non-universal
gaugino masses. We find, as commonly known,  that the relic
density pinpoints towards very specific regions of the  mSUGRA
models: coannihilation, focus and Heavy Higgs annihilation. The
allowed regions widen significantly when varying the top quark
mass. Introducing some non-universality in the gaugino masses
significantly changes this conclusion as in specific non-universal
models the relic density upper limit can be easily satisfied. This
occurs with models where $M_1>M_2$ at the GUT scale when the LSP
has a high wino component.
 Models where $M_3<M_2$
favours the Higgs annihilation channel in large regions of
parameter space and at large $\tan\beta$ also favours the
annihilation of neutralinos in gauge bosons pairs.
 We discuss also the potential of direct detection experiments to probe  supersymmetric models
 and point out at the main consequences  for colliders based on the mass spectrum.
Our calculation of the relic density of neutralinos is based on
\micro~ and the SUSY spectrum is generated with \softsusy.  }
%%%%%%%%%%%%%%%%%%%%%%%%%%%%%%%%%%%%%%%%%%%%%%%%%%%%%%%%%%%%%%%%%%%%%
\vspace*{\fill}

\vspace*{0.1cm} \rightline{LAPTH-1052/04}

\rightline{{\today}}

\end{titlepage}
\baselineskip=18pt
\baselineskip=14pt

Despite the lack of direct  experimental evidence for
supersymmetry, the minimal supersymmetric standard model (MSSM)
remains one of the most attractive extensions of the standard
model. One of  the nice features of the MSSM is that it provides a
natural cold dark matter candidate, the lightest supersymmetric
particle (LSP). Recently, the WMAP satellite has measured
precisely many cosmological parameters, among them the matter
density, the baryonic density as well as the relic neutrino
density \cite{wmap,wmap1}. From these measurements one can infer
the relic density of cold dark matter, $\Omega
h^2=.1126^{+.0161}_{-.0181}$ (at 2$\sigma$). This experimental
measurement severely constrains the parameter space of
supersymmetric models
\cite{Baer:sugra,Ellis:wmap,Nath:wmap,Lahanas:wmap}. This is
particularly true in mSUGRA models where most of the analyses have
been done so far. Although it has been argued that some
realisations of mSUGRA are in fact quite generic to more general
MSSM models so that it is sufficient to restrict to mSUGRA, it
rests that most of the interesting scenarios that might occur in
mSUGRA require a careful adjustment of parameters. For example, the
Higgsino LSP. Moreover some important scenarios are never realised
in mSUGRA like for example having a LSP which is dominantly wino.
Equality of the gaugino masses at the GUT scale ($M'_1=M'_2$)
leads to $M'_1=0.4 M'_2$ at the weak scale thus to a LSP that is
either mostly bino or a mixed bino/Higgsino state. In fact, in
mSUGRA, most of the time, at least in what constitutes from the
relic density point of view the bulk region,  the LSP is an almost
pure bino. The bulk region corresponds to the low $\m0-\mhf$
region of parameter space of mSUGRA models
\cite{Baer:coan,msugra-old,ellisolivesugra}.  Here $\m0(\mhf)$
stands for the common scalar (gaugino) mass defined at the GUT
scale. The recent cosmological data has almost ruled out this
scenario \cite{Ellis:wmap,Nath:wmap,Lahanas:wmap,Baer:chi2}. This
is because the main annihilation channel is into a pair of
fermions and proceeds through the t-channel exchange of a
right-handed slepton, the fermions with the largest hypercharge.
The annihilation of a pure bino is not very efficient since the
U(1) coupling, $g=e/c$,  is small and weaker  than a SU(2) coupling $g=e/s$.
In order to bring down the relic density in the desired range one
must appeal to specific mechanisms, either annihilation of
neutralinos via s-channel Higgs exchange \cite{Drees_heavyhiggs}
or coannihilation of neutralinos with other  sfermions
\cite{Griest:coan}. Finally when the LSP has a significant
Higgsino component, then annihilation into gauge bosons and/or
coannihilation with heavier neutralino/chargino becomes very
efficient \cite{Baer:coan}, so much so that the relic density
tends to be below the range of WMAP \cite{Birkedal-Hansen:2002sx}.

%The main annihilation channel is into a pair of fermions and
%proceeds  through the t-channel exchange of a right-handed
%slepton, the fermions with the largest hypercharge. The
%annihilation of a pure bino is not very efficient since the  U(1)
%coupling, $g=e/c$,  is weaker  than a SU(2) coupling $g=e/s$.
%Annihilation of neutralinos into fermions  can also proceed
%through the exchange of  a Z  in s-channel provide the LSP has
%some Higgsino component. This means $\mu$ small. The amount  of
%necessary Higgsino component depends on how close   the mass of
%the neutralino pair lies in relation to the Z pole. The conditions
%for  neutralinos  annihilation near a Z resonance  as well as
%light sfermions  are met in the low $\m0-\mhf$ region of parameter
%space of mSUGRA models.  Here $\m0(\mhf)$ stands for the common
%scalar (gaugino) mass defined at the GUT scale. However, even in
%the most favourable case it is hard to meet the tight upper bound
%from WMAP \cite{Ellis:wmap,Nath:wmap,Lahanas:wmap,Baer:chi2}. The
%annihilation near a Z resonance is not possible due to direct
%limit from LEP on the chargino mass and the annihilation  with
%sfermion exchange suffers from the small coupling.

 Several analyses have shown that at least one of  these conditions is satisfied only for
 narrow strips in the parameter space of the mSUGRA model \cite{Baer:sugra,Ellis:wmap,Nath:wmap}:
 \begin{itemize}
 \item{}
 coannihilation with $\stauo$   occur in the low $\m0$ region and requires
  $\mstauo \approx \mneuto$ (the coannihilation region),
 \item{}
  annihilation via a  heavy scalar Higgs into a pair of b-quarks
or $\tau$ leptons  is possible at large $\tan\beta$ (the Higgs
funnel region)
\item{}
one finds a neutralino with a high Higgsino content at large values of $\m0$
 when the parameter $\mu$ becomes small ( the  focus point
 region). This occurs near the electroweak symmetry breaking
 (EWSB) border.
 \end{itemize}

 The strong correlation
between the parameters are somewhat responsible for the rather
contrived constraints coming from the relic density in mSUGRA
models. However it was also pointed out that
 when taking into account the uncertainty in the input parameters,
 such as the top quark mass $m_t$, the allowed region widens significantly \cite{Baer:chi2,Ellis:chi2}.
 This is particularly true both  in the focus point region and in the heavy Higgs annihilation region.
In the former,   the value of the $\mu$ parameter, which drives the Higgsino nature of the LSP,  is extremely sensitive to $m_t$ \cite{Romanino,Allanach:codes}. In the latter it is the heavy Higgs masses themselves that depend on $\mt$.
 In the coannihilation region, the top quark mass dependence shows up when applying
  the Higgs mass limit. The direct limit  is somewhat relaxed with a large top quark mass,
therefore  we pay special attention to the $m_t$ dependence. We
confirm qualitatively the results of previous analyses
\cite{Baer:sugra,Ellis:wmap,Nath:wmap} although we differ in the
position of the focus point and of the heavy Higgs funnel regions.

To go beyond the much contrived mSUGRA model while keeping a
manageable number of free parameters, we  introduce
non-universality in the GUT-scale relation for gauginos.
 Such
non-universality can be found in a variety of models that go
beyond mSUGRA. For example, SUGRA models with non-minimal kinetic
terms \cite{nmSUGRA}, superstring models with moduli-dominated or
a mixture of moduli and dilaton fields and anomaly mediated SUSY
breaking models
\cite{nonuniversal-strings,multimoduli,non-universal-anomaly,AMSB,Chen:Drees,Binet:nonuni}
all feature non-universal masses in the gaugino and/or scalar
masses. Here we just introduce two free non-universality
parameters relating the gaugino masses at the high scale,
$\r12=M_1/M_2$ and $r_{32}=M_3/M_2$.

Basically in non-universal gaugino mass models one finds the same
dominant channel for neutralino annihilation, however the
parameter space for which annihilation into leptons, s-channel
annihilation near a resonance, coannihilation or Higgsino/wino
annihilation occurs and give a reasonable value for the relic
density
 differs significantly from mSUGRA model.
 Most importantly, we do not have to carefully tune parameters to fall in the allowed regions.
 It is therefore important to explore the
 relic density as well as other constraints from precision measurements on these categories
 of models in order to assess the potential of both future
 colliders  as well as
 direct/indirect searches  for the LSP. For one,
increasing the ratio $M_1/M_2$ provides a LSP with a much higher
wino  content. Then gaugino annihilation/coannihilation into gauge
bosons becomes the dominant process \cite{Birkedal-Hansen:2001is}.
Values
  for the relic density below the WMAP upper bound are found
 for large regions of the $\m0-\mhf$ parameter space.
 On the other hand decreasing the $M'_1/M'_2$ ratio, one can more easily find light neutralinos
 that can annihilate via a s-channel Z or light Higgs pole
 provided $\mu$ is small enough. This was discussed in details in
 \cite{Belanger:susy02,Belanger:lowneu,Bottino:lowneu}. A shift in  the
 ratio $M'_3/M'_2$ also induces significant changes in the prediction
 for the relic density. Indeed,
 the parameter $M'_3$ has a large impact on the spectrum calculation,
 driving the masses of all coloured particles but also changing the relation between $\mu$ and $M_A$,
 two parameters that are critical for the estimate of the relic density.
 Thus, decreasing $r_{32}$ introduces noticeable shifts in the
 region of parameter space where neutralinos annihilate via a heavy Higgs
 pole. In particular this annihilation channel can be
 relevant at low values of $\tan\beta$.
 Furthermore one finds much lighter squarks than in mSUGRA models favouring
 the  coannihilation with squarks.

Although we focus mainly on the relic density constraint, we also
take into account the experimental constraints set by direct
searches at colliders as well as those coming from precision
measurements, including $\bsgamma$, \gmuon~, $\bsmu$. As concerns
the limit from the muon anomalous magnetic moment,   there have
been many changes  both in the theoretical estimates as well as
improved experimental bounds in the last years \cite{g-2ex}.  In
particular, the standard model prediction  depends on how the
lowest-order hadronic contribution is extracted from the data,
either using $\epem$ or $\tau$ data \cite{Davier:update}. We quote
both the $\epem$ estimate or an average of all results. We think
it would be overly conservative not to include at all this
constraint. After all the upper allowed value for $\delta a_\mu$
has been stable over the many refinements of both theoretical and
experimental results.

When dealing with constraints on the mSUGRA parameters, one
crucial step is the evaluation of the physical spectrum
corresponding to a given mSUGRA model. This is based on the
renormalization group equations (RGE) that specify the evolution
of parameters from the high scale to the low scale as well as on
the inclusion of radiative corrections to the physical masses of
sparticles. This introduces some theoretical uncertainties in the
prediction of the spectrum and can seriously affect the prediction
of observables at low energies and in particular of the relic
density. Here we will present results obtained with \softsusy1.8.6
\cite{Allanach:softsusy}. A  detailed analysis of the difference
between codes for the prediction of the relic density  as well as
a more detailed study of the theoretical uncertainties in the
evaluation of the spectrum and their implications for the relic
density will appear elsewhere
\cite{Allanach:houchesrge,nous_preparation}.

The relic density measurement is not the only way to probe
supersymmetric dark matter in space, many experiments are underway
and even more have been planned to directly search for dark
matter. These experiments are effectively sensitive to either the
spin independent (scalar) or spin dependent neutralino proton
scattering cross section. The neutralino proton cross section
proceeds through Z/Higgs or squark exchange. Thus a large cross
section requires either light squarks and/or significant coupling
with the Z/Higgs. In mSUGRA models this means that large cross
sections are typically found in the focus point region. The
current series of experiments have reached a sensitivity on the
scalar cross section at the level $\sigma_{\neuto-p}\approx
10^{-6}$pb \cite{edelweiss}. The only report of a positive signal
by \dama \cite{dama} has not been confirmed by others {\footnote
{This discrepancy might be explained using different velocity
distributions for light dark matter particles
\cite{GelminiGondoloDama}.}. To  really probe mSUGRA models
however, one still has to wait for improved detectors such as
\cdms2 or \edelweiss II. \cdms2  has recently published much
improved upper limits \cite{cdms2}. By the time one reaches
ton-size detectors, such as  \zepliniv \cite{zeplin}, Genius
\cite{genius} and \xenon \cite{xenon} a signal could be found for
a significant fraction of the parameter space of the mSUGRA model.
At the same time many detectors that are mostly sensitive to the
spin dependent part of the neutralino nucleon cross section have
started taking data, for example \naiad \cite{naiad}, \simple
\cite{simple} and \picasso \cite{picasso}. Upgrades of these
detectors are planned and eventually they will reach a sensitivity
between $10^{-5}-10^{-4}$~pb.

Our calculation of the relic density is based on  \micro1.3
\cite{Belanger:micro13,Belanger:cpc}, a code for the calculation
of the relic density of dark matter
 which includes all annihilation and coannihilation channels. Although all
cross sections are calculated exactly only at tree-level, some
important loop effects are taken into account:
 loop corrections in masses as well as in
vertices, loop corrections  for the Higgs widths are  also
included. The pole masses and all mixing matrices are obtained
from a code that calculates the supersymmetric spectrum.
Additional SUSY corrections to the $hb\bar{b}$ vertices (the
$\Delta m_b$ corrections ) are calculated independently. Although
the radiative corrections to neutralinos, charginos and sleptons
are rather small, they can play a crucial role in the case of
coannihilation. Other constraints from precision measurements are
included as well in the code: $\bsgamma$, \gmuon~, $\bsmu$.

The paper is organized as follows: we first review the model
parameters in section 1 and the direct and indirect constraints in
section 2. In section 3 we  present results in the mSUGRA scenario
taking also into account uncertainties in the top quark mass. We
first summarize the constraints form colliders and precision
measurements before discussing at length the relic density
constraints. As a first step towards a full analysis of the MSSM
we then introduce non-universality in the gaugino sector in
section 4. The potential for  both the
spin-dependent/spin-independent direct detection experiments for
the mSUGRA and non-universal mSUGRA models will be discussed in
Section 5. Our results are summarized in the conclusion.

\section{Model parameters}
\label{model}

The parameters of the MSSM Lagrangian that are necessary for the
calculation of the various processes are obtained using the
renormalization group equations (RGE's) starting with  the input
parameters of mSUGRA models defined at the GUT scale, $\m0$ the
scalar mass, $\mhf$ the common gaugino mass, $A_0$ the common
trilinear coupling,  $\tan\beta$ the ratio of the Higgs vev's
defined at the weak scale  and $sgn(\mu)$, the sign of the Higgs
mixing parameter. The convention used are those of the SUSY Les
Houches Accord \cite{SLHA}, we only repeat here the convention for
the neutralino mass matrix:

\begin{equation}
{\cal M}_{\tilde\chi} \ =\ \left(\begin{array}{cccc} M_1 & 0 &
-\mz \cos\beta \sw & \mz\sin\beta \sw\\ 0 & M_2 & \mz\cos\beta \cw &
-\mz\sin\beta \cw \\ -\mz\cos\beta \sw & \mz\cos\beta \cw& 0 & -\mu \\
\mz\sin\beta \sw & -\mz\sin\beta \cw & -\mu & 0
\end{array} \right)~, \label{eq:mchi0}
\end{equation}
Here $M_i$, the gaugino masses, as well as  other parameters
entering the matrix are defined at the electroweak symmetry
breaking  scale. The nature of the LSP, $\neuto$,  which is a
linear combination of bino $\tilde B$, wino $\tilde{W}$ and the
two Higgsino states $H_{1,2}$,  is a crucial parameter in the
evaluation of the relic density;
 \beqn
\neuto= N_{11}\tilde{B}+ N_{12}  \tilde{W} +N_{13}\tilde{H_1}+
N_{14}\tilde{H_2}
 \eeqn
 where $N$ is the neutralino mixing matrix. By inspection of the neutralino matrix one sees that
 if $M_1<<M_2,\mu$ the lightest  neutralino is almost a pure bino,
 if $\mu<<M_1,M_2$ it becomes dominantly Higgsino and when $M_2<<M_1,\mu$
 the neutralino is dominantly wino.
 In mSUGRA models, one always
finds that $M_1\approx 0.4 M_2$ at the electroweak scale, moreover
in most cases $M_2<\mu$.
 Thus the lightest neutralino is mainly a bino. This
 also leads to the usual mass relation between the LSP and the lightest chargino
  $\mneuto\approx .5\mchargo$. Within the mSUGRA model,
there also exists a possibility for a LSP with a much larger
Higgsino component, this occurs  in the focus point region at
large $\m0$.
 There, the parameter $\mu$ decreases very rapidly until  eventually $\mu^2<0$ and
 the EWSB cannot take place.  When $\mu< M_1,M_2$,  the mass of
 the LSP is driven by $\mu$ and one has $\mneuto\approx\mneutt\approx\mchargo$.
 The coupling of the neutralino to  the Z depends on the  Higgsino fraction
 and is therefore enhanced in the focus point region. The coupling
 of the neutralino to the light Higgs also requires some Higgsino
 component and is therefore also enhanced in the focus point region of mSUGRA.

The parameter $\mu$ is extracted from the  minimization of the
potential at the electroweak scale.  At tree-level, \beqn
\mu^2=\frac{m_{H_1}^2-m_{H_2}^2\tan^2\beta}{\tan^2\beta-1}-\frac{1}{2}\mz^2
\eeqn where $m_{H_1},m_{H_2}$ are the two soft scalar masses in
the Higgs potential. Thus the value for $\mu$ strongly depends on
the RG evolution of these scalar masses. In mSUGRA one finds
typically
 large values for $\mu$ as compared to the gaugino masses.
  In  more general  models defined at the GUT scale
 one introduces new parameters at the GUT scale which can change the  relation among parameters
 at the weak scale,  it is then possible to obtain a LSP with a significant wino or Higgsino component.

In models with  non-universal gaugino masses we introduce two
extra parameters that characterize the amount of non-universality,
\beqn r_{12}=\frac{M_1|_{GUT}}{M_2|_{GUT}}\;\;\;\;\;\;
r_{32}=\frac{M_3|_{GUT}}{M_2|_{GUT}} \eeqn
 where $M_i|_{GUT}=M_i'$ are the gaugino masses
defined at the GUT scale. We always  define $M'_2=\mhf$.

Relaxing the gaugino masses universality, one can have models
where the LSP can have a large wino component. For example,
$r_{12}>1$, leads to    a LSP that has a significant wino
component. LEP data then implies that $\mneuto> M_W$ which in turn
 favours large annihilation cross sections for neutralinos into
gauge bosons. This also implies that the chargino can be almost
degenerate with the LSP, $\mneuto\approx\mchargo$, hence the
important contribution of coannihilation channels.

 The parameter $r_{32}$ has a strong  influence on the weak scale parameters.
 Not only does the gaugino mass $M_3$ determine the coloured
sector but it  also  enters indirectly the RGE equations for
$M_{H_1},M_{H_2}$. In particular, this implies that  both $\mu$
and $\ma$ also depend on $M_3$ and can be significantly shifted as
compared to the unified case. Taking all scalar masses equal at
the GUT scale and with $\tan\beta=10$, $A_0=0$, we obtain an
approximate solution for $\mu^2$ at the weak scale,  \beqn
\label{mu} \mu^2&\approx& -0.5 \mz^2 + c_0 \m0^2+2.6
{M'}_3^2-0.2{M'}_2^2+0.005
{M'}_1^2+0.034 M'_3M'_1\nonumber\\
&&+ 0.22M'_3M'_2+0.007 M'_2M'_1 \eeqn
 The coefficient $c_0$
depends critically on the top mass at the SUSY scale, for
$m_t=175$~GeV, $c_0\approx 0.11$. $c_0$ can get very small and
even negative which defines the focus point regions where $\mu$
becomes very small.
 Clearly then because of the partial cancellation between the
 term in $M'_3$  and the one in $M'_2$, when $r_{32}<1$, $\mu$ is
 smaller than in the unified case. This once again leads
 to a LSP  with a higher Higgsino component and more
degenerate with the lightest chargino.  The former implies more
efficient annihilation, the latter more efficient coannihilation.

 The pseudoscalar mass also depends strongly on $M_3$. An
approximate solution to the RGE equations under the same condition
as above leads to \beqn \label{ma} \ma^2\approx -M_Z^2+1.08
M_0^2+2.6{M'}_3^2+0.28{M'}_2^2+0.22M'_2M'_3+0.03{M'}_1^2.\eeqn
When $r_{32}<1$,  the pseudoscalar mass is lowered as compared to
the universal case and one can find a pole in the neutralino
annihilation channel even for an intermediate value of
$\tan\beta$. For example, when $r_{32}=1/2$ , $\mneuto^2=.4M_2^2$
and $\m0\approx .7M_2$, we find $\ma\approx 2\mneuto$. In the
universal case such a solution can be found only at large values of
$\tan\beta$. This will open up new regions of parameter space
where neutralino annihilation proceeds through an exchange of a
Higgs in the s-channel.

Introducing non-universal Higgs masses would in many ways
reproduce  the features of the  non-universal gaugino masses
scenarios in the sense that both $\mu$ and $\ma$ are directly
related to the Higgs scalar masses at the EWSB scale. For example,
lowering $M_3$ is equivalent to taking $M_{H_1}$ larger than other
scalar masses. Then it is possible to obtain values for $\mu$ and
$\ma$ that are lower than in the universal case leading to more
efficient annihilation/coannihilation of neutralinos.
 We do not discuss
these scenarios any further, they have been discussed in
\cite{Ellis:nonuni_higgs}.

In this paper we do not use the approximate solutions
Eq.~\ref{mu}-\ref{ma} which only serve as a guide, but obtain the
electroweak scale parameters directly by solving completely the
RGE's. This is done by a call to \softsusy, a program that solves
the RGE equations and also calculates the radiatively corrected
superparticle masses and couplings.  These higher-order
corrections  are essential to precisely evaluate the relic density
both in the coannihilation regions where the exact mass splitting
is a crucial parameter and near a s-channel resonance.

\section{Constraints on
supersymmetric models: Direct limits, $\Omega h^2$, \gmuon, $b\ra
s\gamma$, $B_s \ra \mu^+ \mu^-$}

\noi
{\bf Direct limits from LEP}

For the chargino mass, we use the LEP2 bound, $\mchi> 103.5$GeV except
when there is near degeneracy with the lightest neutralino
($\Delta M ={\cal O} (1)$GeV), then $\mchi> 91.9$GeV
\cite{lep_susy_WG}.

For the neutralino sector we do not use directly the lower limit
on the neutralino mass quoted by the LEP experiments, $\mneuto
<59.6GeV$ since this value implicitly assumes  universal gaugino
masses   at the GUT scale, $M_1=M_2$ \cite{lep_neutralino}.
Rather, we use in addition to the constraint on the chargino mass,
the processes $\epem\ra \chi_1^0\chi_2^0, \chi_1^0\chi_3^0$. The
associated production can somewhat improve on the constraints set
from the chargino pair production in scenarios where $M_1<M_2$
\cite{Belanger:hinvisible,Belanger:lowneu} in particular in the
region with small $\mu$. In our scans, we implemented the upper
limit from the L3 experiment on
$\sigma(\epem\ra\neuto\neutt+\neuto\neutth\ra\slashE \mu^+\mu^-)$
\cite{L3_susy}. The radiative process $\epem\ra\neuto\neuto\gamma$
and the invisible width of the Z, $\Gamma_{Z_{inv}}<3MeV$,  can in
principle also constrain the neutralino sector but these can play
a role only in models with light neutralinos
\cite{Belanger:lowneu}. Even then, present limits are such that
these processes do not help much in reducing the allowed parameter
space.

For selectrons, a limit of $\mser>99.4$~GeV can be set when
$\mneuto>40$GeV while for  other charged leptons the lower limits
of
 $m_{\smuon} > 96.5$GeV  and  $m_{\stau} > 92.5$GeV
apply \cite{lep_susy_WG}.

In  models where the pseudoscalar mass is heavy, the limit on the
Higgs mass from LEP2, $m_H>114.4$ GeV, applies. However we have
imposed the limit $m_H>111$ GeV to allow for theoretical
uncertainties. The Higgs mass is calculated by the  code for the
SUSY spectrum calculation, here \softsusy1.8.6,  and includes the
two-loop corrections of Feynhiggs \cite{Heinemeyer:cpc} as well as
the two-loop yukawa corrections \cite{slavich}. In models with a
light pseudoscalar the above LEP2 constraint is relaxed,
% . When
%$m_h\approx M_A$ and $\cos(\alpha-\beta)\approx 1$, the channel
%$\epem\ra hZ$ is strongly suppressed and LEP2 can only make use of
%the $hA\ra bbbb,\tau\tau b b$ channels. We have used the
%$\cos(\alpha-\beta)$ dependent bound on $m_h$  with
 the absolute bound  $m_H,M_A> 91.6$~GeV is obtained when
$\sin(\alpha-\beta)\approx 0$ \cite{aleph_HA,combined_HA}.

\newpage

\noi {\bf Relic density of neutralinos $\Omega h^2$}

The supersymmetric models with a stable neutralino
 must be consistent with at least  the upper limit on the amount of cold dark matter.
 We consider both the $2\sigma$ limit from WMAP,
$$0.094< \Omega h^2< 0.129$$
 as well as only the upper limit since there could be some
other contribution to the cold dark matter. Our calculations of the
relic density is based on {\micro}1.3, a program that calculates
the relic density in the MSSM including all possible
coannihilation channels \cite{Belanger:cpc,Belanger:micro13}.
Within the context of a model defined at the GUT scale such as
universal/non-universal SUGRA
 that we consider here, we first need to calculate the MSSM parameters.
 To do this we use a renormalization group evolution code,
 \softsusy~
to evaluate the supersymmetric mass spectrum as well as the mixing
matrix elements
  following the convention of the SUSY Les Houches Accord
 \cite{SLHA}. These parameters
 are then used  together with
 $\tan\beta$ and $\mu$     to calculate all cross sections for annihilation
 of neutralinos or coannihilation. Although the mass and mixing parameters include
  one-loop corrections,
this procedure is of course   not a complete one-loop computation
of the relic density.    However the most important one-loop
effects are taken into account. In particular the  mass difference
between the NLSP and the LSP are computed correctly.  This mass
difference is crucial in evaluating the relic density in the
coannihilation region. Indeed in computing the effective
annihilation cross section, coannihilation processes are
suppressed by a Boltzmann factor $\propto \exp^{-\Delta M/T_f}$
where $\Delta M$ is the mass difference between the NLSP and the
LSP and $T_f$ the decoupling temperature ($T_f\approx
M_{LSP}/25$). Then   even a small shift in the masses can reflect
significantly on the relic density.  For the Higgs width the
running of the b-quark mass including three loop corrections as
well as the $\Delta m_b$ corrections are included
\cite{Spira:dmb}.

\noi{\bf Muon anomalous magnetic moment  $\mathbf{(g-2)_\mu}$}

The latest experimental data on the $g-2$ measurement using
$\mu^-$\cite{g-2ex}, brings the average to \beqn a_\mu^{\rm
exp.}=11659208 \pm 6 \times 10^{-10} \eeqn

The quantity $a_\mu$ includes both electroweak and hadronic
contributions and is still subject to large theoretical errors,
$$a_\mu^{\rm theo.}=a_\mu^{\rm QED}+a_\mu^{\rm weak}+a_\mu^{\rm
HAD}+a_\mu^{\rm LBL}+a_\mu^{\rm HAD/NLO}.$$ The largest
uncertainty in the calculation of $a_\mu$ arises from the lowest
order hadronic contribution. We take an average of four recent
estimates: the value reported by Jegerlehner $a_\mu^{had}=(6889\pm
58)\times 10^{-11}$ \cite{Jegerlehner:g-2} that includes BES
and CMD-2,
 the value extracted from inclusive data by Teubner  {\it et al.}
$a_\mu^{had}=(6831\pm 59\pm 20)\times 10^{-11}$
\cite{Teubner:2002},  the latest estimate by Davier {\it et al.} based on
\epemt data $a_\mu^{had}=(6963\pm 62_{exp}\pm 36_{th})\times
10^{-11}$\cite{Davier:update} and finally the result of Yndurain,
$a_\mu^{had}=(6935\pm 50_{exp}\pm 10\pm 5_{th})\times
10^{-11}$\cite{Yndurain:2004g2}. The latter two authors also
report  another estimate that is based on $\tau^+\tau^-$ data. The
values extracted from $\tau$ data are consistent with each other
but differ from the values obtained from $\epem$
\cite{Davier:update}. For the  hadronic light by light
contribution we take $a_\mu^{LBL}=(+80\pm 40)\;\times 10^{-11}$
\cite{Nyffeler:g2} which is based on the calculation of
Ref.~\cite{Knecht:2002hr,Knecht:2001qg,Knecht:2001qf} and includes a
large theoretical error.

We get, after  averaging the different hadronic contributions from
$e^+e^-$ data alone, \beqn \delta a_\mu=a_\mu^{\rm exp.} -
a_\mu^{\rm theo.}=\left(33.8\pm 6_{|{\rm exp.}} \pm 11.5_{|{\rm
theo.}}\right) \;\times 10^{-10} \eeqn while including also the
$\tau$ estimate we get \beqn \delta a_\mu=a_\mu^{\rm exp.} -
a_\mu^{\rm theo.}=\left(28.5\pm 6_{|{\rm exp.}} \pm 11.5_{|{\rm
theo.}}\right) \;\times 10^{-10} \eeqn
 Adding the errors in quadrature leads to the
$2\sigma$ allowed range, for the
$e^+e^-$ data alone,  \beqn 7.8 \;<\; \delta a_\mu\;\times 10^{10}
\;<\; 59.7 \label{3sigma_1} \eeqn while including $\tau$ data in
the average,  the value moves closer to the standard model, \beqn
2.5 \;<\; \delta a_\mu\;\times 10^{10} \;<\; 54.4
\label{3sigma_2} \eeqn

Note that our central value is slightly higher than the one
reported by Davier~\cite{Davier:update} as we have averaged all
estimates of the lowest-order hadronic contribution. We will in
general only show contour levels for $\amu$. We remark that  in
the last couple of years there has been many improvements  to the
computation of the standard contribution to \gmuon, and the
allowed area has shifted significantly.  Still the upper bound has been rather stable and
therefore  should be considered as a conservative constraint on
the supersymmetric model.

\noi{ $\mathbf{\bsgamma}$}

Our calculation of the branching ratio for $b\ra s\gamma$ takes
into account next-to-leading (NLO) order terms, bremstrahlung and
some non-perturbative effects \cite{Kagan:bsg,bsgCMM,GambinoMisiak:bsg}.  Our
standard model value (with scale parameters set at $m_b$), gives
$Br(b\ra s\gamma)=3.72 \;\times 10^{-4}$ while the scale and other
parameter uncertainty ($\alpha_s$, CKM ...) are about
$10\%$ \cite{Belanger:micro13}. The Wilson coefficients for the
charged Higgs contribution are evaluated at
NLO\cite{Ciuchini:smh} while SUSY contributions at
leading-order are included as well as the important large $\tgb$
effects \cite{bsgDGG}

The experimental  value for the branching ratio is extracted from an average of
 the Babar\cite{Babar:bsg}
    CLEO \cite{CLEO:bsg} BELLE\cite{bsgBELLE} and
ALEPH \cite{bsgALEPH} measurements \beqn \label{bsgexp} Br(b\ra
s\gamma)=3.34\pm .38 \;\times 10^{-4} \eeqn and assumes a fully
correlated theoretical error\cite{Jessop:2002}.

We require that after allowing for the (scale) uncertainty in the
theory calculation the result must be within $2\sigma$ of the
experimental result, Eq.~\ref{bsgexp}. Since the theory
uncertainty is roughly constant over the SUSY parameter space we
have allowed for  a conservative fixed uncertainty of $10\%$
independently of the SUSY parameters. Thus in effect we require
the theory prediction to fall within the range \beqn
\label{bsgbound} 2.25 < Br(b\ra s\gamma)\;\times 10^{4} < 4.43
\eeqn

\noi{\bf $\mathbf{\bsmu}$}

The CDF experiment at Fermilab has obtained an upper bound on the
branching ratio $Br(\bsmu)<9\times 10^{-6}$ \cite{CDFbsmumu} and
should be able to reach $Br(\bsmu)<2\times 10^{-7}$ in RunII. In
the SM, this branching ratio is expected to be very small
($\approx 3\times 10^{-9}$). In the MSSM, SUSY loop contributions
due to chargino, sneutrino, stop and Higgs exchange can
significantly increase this branching ratio. In particular, the
amplitude for Higgs mediated decays goes as $\tan\beta^3$ and
orders of magnitude increase above the SM value  are expected for
large $\tan\beta$.  Our calculation is based on
\cite{bsmumubobeth} and agrees with \cite{bsmumudreiner}. The
$\Delta m_b$ effects relevant for high $\tan \beta$ are taken into
account.

\section{Results in mSUGRA}
\label{results}

We have performed scans over the mSUGRA parameter space using the
code  \softsusy1.8.6. We first present results in the $\m0-\mhf$
plane for different choices of  $A_0$ and $\tan\beta$ and with
fixed values of the top quark mass. We discuss constraints from
precision measurements before concentrating on the relic density
constraints. We consider only  models with $\mu>0$ as they are the
ones favoured by the \gmuon~ measurement \cite{g-2}. We note from
the onset that there are theoretical uncertainties in the
evaluation of the spectrum of the MSSM within mSUGRA models
 \cite{Allanach:codes}.
 These can have a large impact on the predictions for the relic density \cite{Allanach:houchesrge}.
 Here we fix $\mbmb=4.23$~GeV and choose values of $\mt$ within the $1\sigma$ range,
 $\mt=174.3\pm 5.1$~GeV \cite{PDG:2002}.
 {\footnote {New results from  the Tevatron indicate a higher value for the top quark mass $\mt=178\pm4.3$\cite{top_tevatron}. The new central  value  falls within the range we have considered.}}

\subsection{Precision measurements and direct limits}

The mass of the lightest  Higgs  is basically set by the top and
the stop quark sectors. Heavier masses are found in regions with
large $\tan\beta$ and large mixing in the stop sector.
 One expects a sensitivity to
the input value of $m_t$, as well as to the value of $A_0$ which
sets the parameter $A_t$ that drives the mixing among the stop
quarks. Large mass splitting between the stops  is difficult to
find in the low $\m0-\mhf$ region of parameter space. In this
region the Higgs tends to be too light as seen from the contours
of constant Higgs masses $\mh=111, 114$ GeV in
Fig.~\ref{mh_contour}. An increase in the top quark mass also
increases the Higgs mass. This has particularly a large impact on
models with small values of $\tan\beta$. For example, for
$\tan\beta=5(10)$ and for small $\m0$, the allowed region shifts
from $\mhf>417(271)$ GeV for $m_t=175$ GeV to $\mhf>332(230)$~GeV
for $m_t=179$ GeV. For $\tan\beta=10$, the Higgs mass limit alone
translates into a lower bound on the LSP $\mneuto> 105(87.4)$ GeV
as well as on the lightest chargino $\mchargo> 194(160)$ GeV
respectively. Of course lighter charginos, in fact down to the
direct limit set by LEP, are allowed for larger $\m0$ which
correspond to  heavier sleptons, but these will in general not be
relevant from the relic density point of view as we will discuss
next.  A decrease in the top quark mass conversely tightens the
constraint on $\mhf$, for example the $m_t=172$~GeV and
$\mh=111$~GeV contour roughly coincides with the $m_t=179$~GeV,
$\mh=114$~GeV contour of Fig.~\ref{mh_contour}a. At large
$\tan\beta$, the Higgs mass increases as well, so that the
constraint on both $M_0$ and $\mhf$ is relaxed.

\begin{figure*}[htbp]
\begin{center}
\vspace{-1.2cm}
\includegraphics[width=16cm,height=10cm]{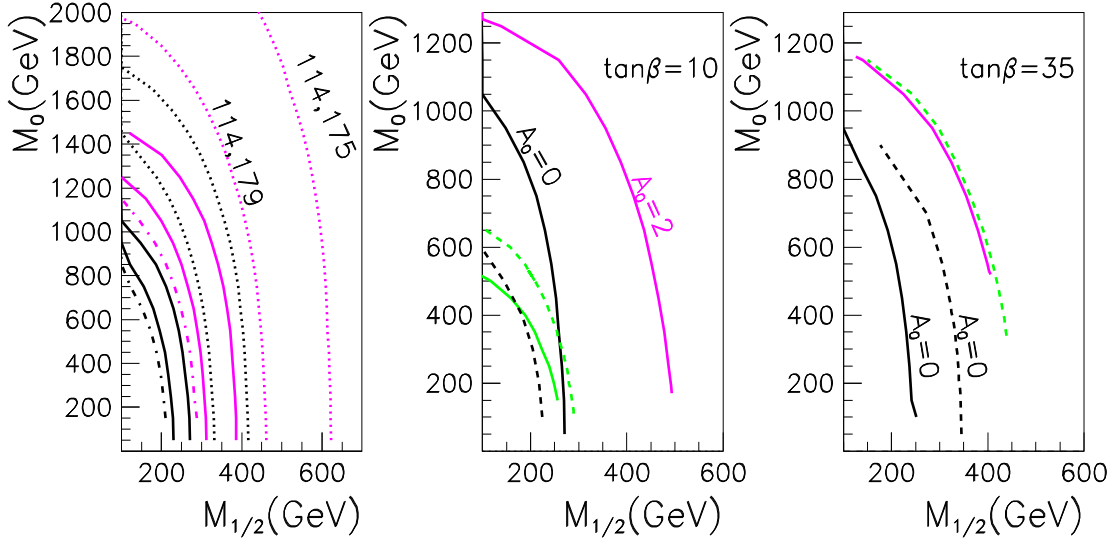}
\vspace{-1.9cm} \caption{\label{mh_contour}{\em  a) Contours of
$m_h=111$~GeV (black) and $m_h=114$~GeV (pink/grey) in the
$\m0-\mhf$ plane for $\mu>0$, $A_0=0$  and for $\tan\beta=5$
(dot), $\tan\beta=10$ (full), $\tan\beta=35$ (dot-dash). Both
contours for $\mt=175$~GeV and $\mt=179$~GeV are displayed. For
each set of parameters, the contour for the heavier $\mt$ is on
the left. For $\tan\beta=35$, only the contour for $\mt=175$~GeV
is displayed. Some $\mh,\mt$ values are attached to the contours
for easy reference. b) Effect of the trilinear mixing for
$\tan\beta=10$, $A_0=0$ (black) $A_0=-1$TeV (green/light grey),
$A_0=2$TeV (pink) on the $\mh=111$~GeV contours (full) and
$Br(\bsgamma)=2.25\;10^{-4}$ contours (dash). c) Contours for
$\mh=111$GeV (full),  $\tan\beta=35$ with $A_0=0$ (black),
$A_0=2$TeV (pink) and for $Br(\bsgamma)=2.25\;10^{-4}$ (dash) with
$A_0=0$ (black), $A_0=-1$TeV (green/light grey). }} \vspace{-.2cm}
\end{center}
\end{figure*}

The shift in the  Higgs mass contours as one varies $|A_0|< 2000$
GeV is quite significant as displayed in Fig.~\ref{mh_contour}b.
First note that imposing $A_0=0$ in mSUGRA, leads to a negative
value for $A_t$ at the electroweak scale. Since the dominant
correction to the Higgs mass depends on the mixing in the stop
quark sector,   $\propto A_t+\mu/\tan\beta$, larger Higgs masses
will be found for $A_0<0$ where $|A_t|$ is large. Indeed in
Fig.~\ref{mh_contour}b, one sees that with $A_0=-1$TeV, the Higgs
mass increases loosening the constraints on the low $\m0-\mhf$
corner of parameter space for $\tan\beta=10$. Already at
$\tan\beta=35$, one finds $m_h>111$~GeV over the full parameter
space. Conversely, constraints are tightened for $A_0=2$~TeV. Note
that the region where one can find a consistent solution to the
RGE's  depends on $\mt, \tan\beta$ and on the value of $A_0$.
Hence some of the contours stop abruptly in Fig.\ref{mh_contour}
when one meets a non-physical region. Furthermore, the region
where the LSP is neutral also shifts with $\tan\beta, m_t$ and/or
$A_0$ mainly because of the mixing in the $\stau$ sector as will
be discussed in the next section.

The rare processes $\bsgamma,\bsmu$ as well as $\amu$  all
provide large supersymmetric corrections at large values of
$\tan\beta$ especially when the sfermions are light. The branching
ratio for $\bsgamma$ depends mostly on the squark and
gaugino/higgsino sector as well as on the charged Higgs. Even
heavy squarks do not completely decouple and one can get
substantial corrections to the SM branching ratio. The chargino
exchange diagram gives a negative contribution relative to the SM
one (for $\mu>0$) and the branching ratio for $\bsgamma$ drops
below the allowed band in the low $\m0-\mhf$ region. Furthermore
there is a strong $A_t$ dependence  from the
mixing in the stop sector.
 For $\tan\beta=10$ and $\mt=175$~GeV,  the $\bsgamma$ constraint is not as severe as the
constraint from the Higgs mass as long as $A_0\geq 0$,
Fig.~\ref{mh_contour}. For $A_0=2$~TeV, the $\bsgamma$ constraint
is satisfied over the full parameter space. For a large negative
value for the trilinear coupling, $A_0=-1$~TeV,  the $\bsgamma$
constraint dominates. As one increases $\tan\beta$, for example
$\tan\beta=35$, the dependence on $A_0$ is even more striking both
for the Higgs mass and for the $\bsgamma$ branching ratio,
especially that the shifts on the allowed parameter space work in
opposite directions. For $A_0=2$~TeV, the Higgs mass constraint is
the most severe since the $\bsgamma$ plays no role while for
$A_0=-1$~TeV,  the $\bsgamma$ constrains
 values of $\mhf$ all the way
to 450GeV whereas the Higgs mass constraint has no impact, see
Fig. \ref{mh_contour}b.
 For  $A_0=0$, it is the  $\bsgamma$ constraint that dominates.
 In the region of light sleptons which will be relevant from the
relic density point of view, the lower bound on $\mhf=
345(435)$~GeV for $\tan\beta=35$ and $A_0=0 (-1)$~TeV implies
$\mneuto>139 (181)$~GeV and $\mchargo>261(347)$GeV.
 As one
further increases the value of $\tan\beta$ one retains the same
features: only the size of the forbidden region at low $\m0-\mhf$
increases.

The branching ratio for $\bsmu$ is predicted to be much below
current experimental limits at low $\tan\beta$, however it
increases rapidly with $\tan\beta$.  There the chargino/stop
exchange contribution dominates and it is largest  in the low
$\m0-\mhf$ region\cite{Arnowitt:bsmu}. This constraint depends somewhat on
the top quark mass which influences the spectrum in the stop
sector. The branching ratio is also dependent on the value of the
mixing in the stop sector. Negative values of $A_0$, which as we
have discussed above decrease $|A_t|$, strengthen the constraint.
 Eventually $\bsmu$ will become
an important constraint, this occurs  typically for larger values
of $\tan\beta$  that we have considered \cite{Arnowitt:bsmu}.
Among the cases studied here   it is only for $\tan\beta=50$ that
in RunII at Tevatron one could put restrictions on the allowed
parameter space, for example for the case $A_0=-1000$~GeV  shown
in Fig.\ref{a-1000}.

 As for \gmuon, the allowed parameter
space depends critically on how much of a deviation from the
standard model one imposes. A significant contribution to $\amu$
is obtained with  light charginos and light sleptons. The upper
limit on $\delta a_\mu$ then rules out the lower left hand corner
of the $\m0-\mhf$ plane. Larger deviations from the standard model
are predicted as one increases $\tan\beta$
%leading to a larger exclusion area (for $\mu>0$)
   but this constraint does not play
an  important role once the limit on the Higgs mass is taken into
account. On the other hand, requiring a non zero deviation from
the standard model implies not so heavy sleptons and neutralinos,
thus constraining the large $\m0-\mhf$ region of parameter space
as well . This region is more important at  small values of
$\tan\beta$ where, as just
 mentioned, one tends to get predictions consistent with the standard
 model, as displayed in Fig.\ref{sugra_mt175} for the contour $\amu=5.\times 10^{-10}$.
 Introducing a
negative value for  $A_0$ does not have a large impact on  the
prediction for $\amu$ as it is the combination
$A_\mu+\mu\tan\beta$ that comes into play in the smuon mixing. For
intermediate to large $\tan\beta$, the mixing is dominated by
$\mu\tan\beta$.

To summarize the collider constraints in mSUGRA type models, the
mass of the Higgs is the main constraint in the low $\m0-\mhf$
region for  values of $\tan\beta< 35$ and is dependent both on the
value of the top quark mass and the trilinear mixing. For
$\tan\beta\geq 35$, the $\bsgamma$ constraint takes over, at least
for $A_0$ not too large or for $A_0<0$.
%For example  the $\m0< 1050(950)$~GeV, $\mhf<260(350)$ GeV region
%is strictly ruled out for $\tan\beta=10(35)$, $\mt=175$ GeV and
%$A_0=0$. This increases to $\m0> 1250(1180)$~GeV, $\mhf>490(440)$ GeV
%respectively for $A_0=2$TeV.
 If one takes into account  the lower
limit from $\amu$  one ends up with an allowed band in the
$\m0-\mhf$ plane. This band moves towards higher values of $\mhf$
and increases in size with increasing $\tan\beta$. We stress again
that the upper limit on $\amu$ is rather conservative but the
lower limit is still subject to debate. In the following we will
display the contours for fixed values of $\amu$.

\subsection{Relic density}

Within the context of mSUGRA models, one needs a special tuning of
parameters to be in  agreement with the very precise measurement
on the relic density by WMAP.   We first remark that the so-called
bulk region (at low $\mhf$ and $\m0$) that used to be one of the
favoured regions \cite{Baer:coan,msugra-old} has considerably shrunk.
  In the bulk region the LSP is almost a pure bino. As such,  the LSP couples mostly to
  fermion/sfermions with the largest hypercharge, that is the right-handed sleptons.
The dominant
annihilation channel for neutralino annihilation  is into  a pair of fermions via the
  t-channel exchange of a right-handed sfermion.  This annihilation process
is not efficient enough to satisfy the new tight upper limit on
the relic density of dark matter.  One needs some additional
contribution from the $\stauo$-coannihilation or Higgs exchange channels.
In fact, the  region at  low $\m0-\mhf$ is only allowed at large
$\tan\beta$
 because there one gets extra contribution from s-channel Higgs
 exchange as well as from coannihilation.
As we have just discussed  the low $\m0-\mhf$ region is also
constrained from the Higgs mass limit
  at least for intermediate $\tan\beta$ and by
 the measurement of $\bsgamma$ for  $\tan\beta>35$.
We also remark  that the top quark mass plays an important role in
determining the relic density
 constraint in mSUGRA models. In some cases
the special tuning of parameters required to satisfy the relic
density constraint can only be obtained for specific values of the
top quark mass. For example
 two of the preferred regions, the focus point region and the
 heavy Higgs funnel regions,  are very sensitive, through the
 renormalization group equations, to the value of the running top
 quarks mass. We first discuss in detail the case $\mt=175$~GeV
 before letting the mass vary within the $1\sigma$ range.

\subsubsection{$\mt = 175$~GeV}

 For any value of $\tan\beta$, the
regions consistent with the WMAP measurement correspond, for a
fixed value of $m_t$, to thin strips in parameter space. The WMAP
allowed region corresponds to the very narrow area between the
contours $\Omega h^2=.094$ and $\Omega h^2=.129$,   see for
example Fig.~\ref{sugra_mt175} for $m_t=175$~GeV.  We are then
left mainly with  three regions: $\stauo$-coannihilation, focus point and
Higgs funnel. Although we will generally refer to the Higgs funnel
region as the region where neutralino annihilation proceeds
through the s-channel exchange of a heavy Higgs scalar, we note
that there is also a small region where light neutralinos
annihilate efficiently near the light Higgs resonance. For
example, this can be seen as the narrow vertical strip along the
LEP2 limit on the chargino mass near $\mhf=150$~GeV, in
Fig.~\ref{sugra_mt175} for $\tan\beta=10$.

%Annihilation of neutralinos into fermions  can also proceed
%through the exchange of  a Z  in s-channel provide the LSP has
%some Higgsino component. This means $\mu$ small. The amount  of
%necessary Higgsino component depends on how close   the mass of
%the neutralino pair lies in relation to the Z pole. The conditions
%for  neutralinos  annihilation near a Z resonance  as well as
%light sfermions  are met in the low $\m0-\mhf$ region of parameter
%space of mSUGRA models.  However, even in
%the most favourable case it is hard to meet the tight upper bound
%from WMAP \cite{Ellis:wmap,Nath:wmap,Lahanas:wmap,Baer:chi2}. The
%annihilation near a Z resonance is not possible due to direct
%limit from LEP on the chargino mass and the annihilation  with
%sfermion exchange suffers from the small coupling.

\begin{figure*}[tbhp]
\begin{center}
%\vspace{-1.2cm}
\includegraphics[width=16cm,height=10cm]{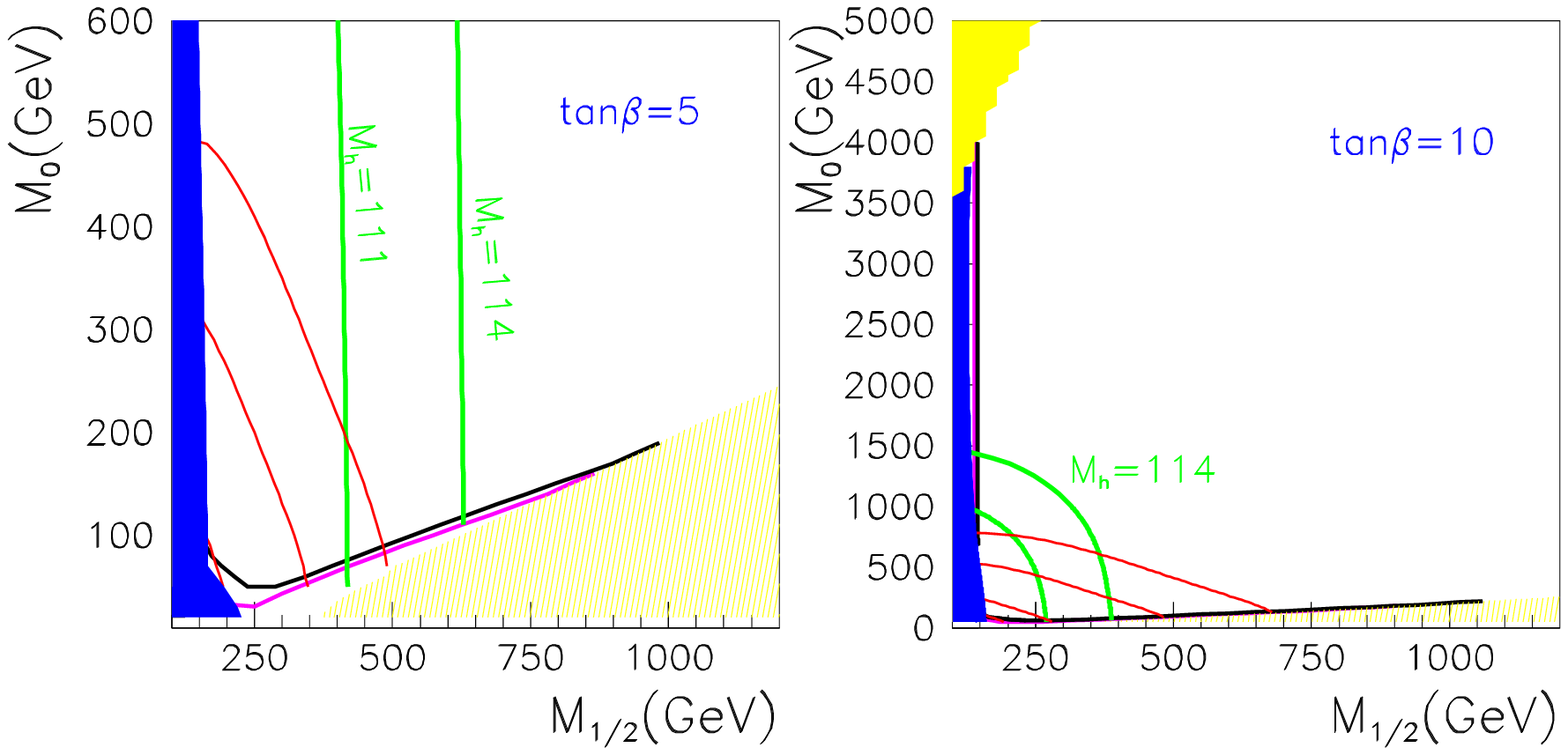}
\includegraphics[width=16cm,height=10cm]{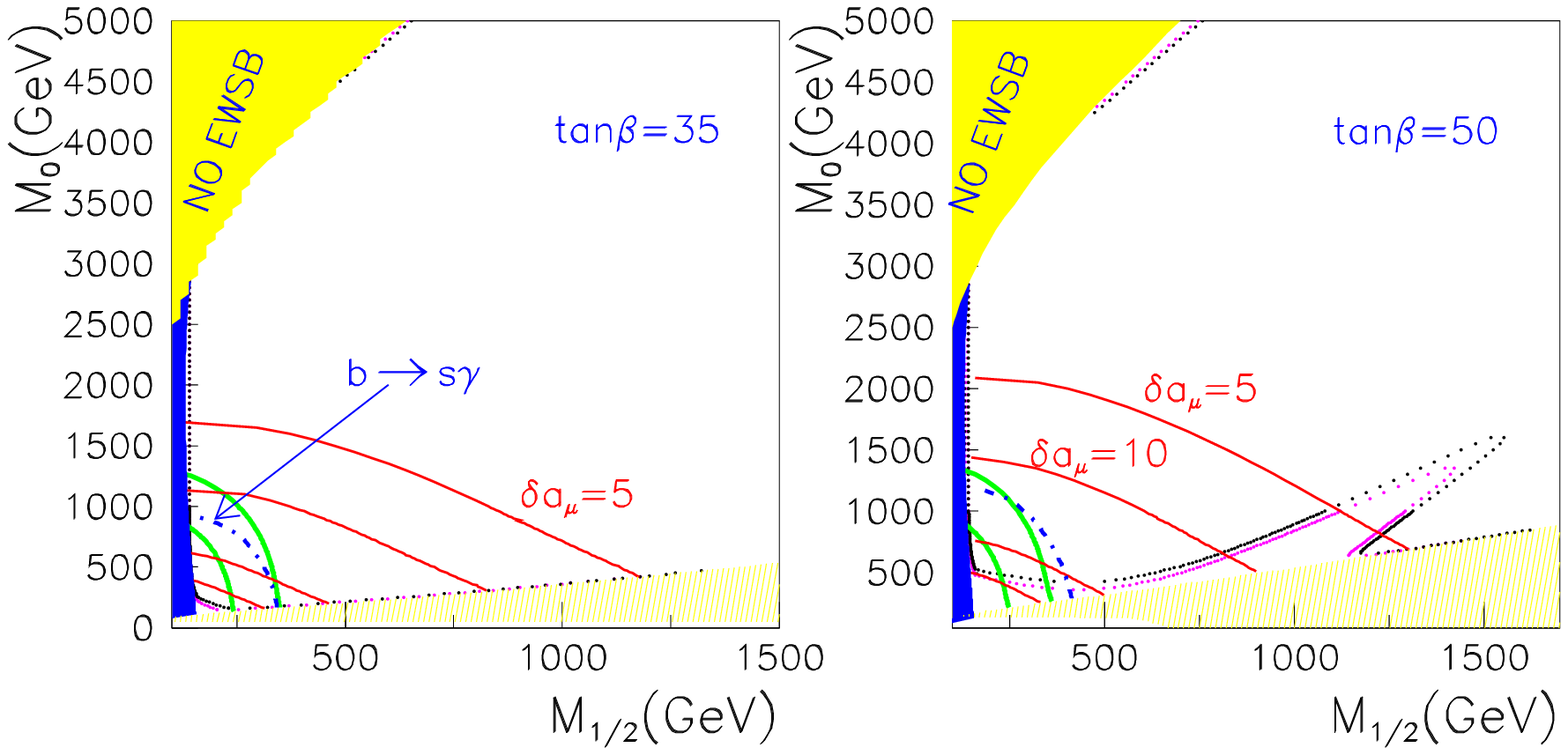}
\vspace{-1.9cm} \caption{\label{sugra_mt175}{\em Allowed regions
in the $\m0-\mhf$ plane  for $\tan\beta=5,10,35,50$, $A_0=0$,
$\mu>0$ and $\mt=175$~GeV. Contours for $\mh=111,114$GeV (green)
$\amu=5,10,30,60\times 10^{-10}$ (red) (top to bottom) and for $\Omega h^2= .129 (.094)$ (black/pink dots). The
contour $\bsgamma=2.25 \times 10^{-4}$ (dash-dot/blue) is also
displayed for $\tan\beta=35,50$.  The region to the left of this
contour is excluded. The LEP direct limits exclude the region
delimitated by the blue vertical band. The regions between the
$\Omega h^2=.094$ contour and  the theoretically excluded regions
(yellow/light grey) have a value for the relic density below the WMAP bound.
}} \vspace{-.2cm}
\end{center}
\end{figure*}

In mSUGRA, the $\stauo$-coannihilation region designates the region at
 low $\m0$, where the lightest $\stauo$ is the NLSP. As the relic density
is very sensitive to the NLSP-LSP mass difference, the proper mass
degeneracy for a relic density consistent with WMAP corresponds to
a very narrow strip in the $\m0-\mhf$ plane, see
Fig.~\ref{sugra_mt175}. Naturally this strip lies slightly above
 the forbidden region where $\mstauo<\mneuto$. This
strong dependence on the mass difference also means that it is
crucial to properly determine the pole masses from the running
$\overline{DR}$ masses. In the coannihilation region, the main
channels are $\neuto\stauo\ra \tau \gamma$ and $\stauo\stauo\ra
\tau\bar\tau$. As one moves towards degenerate $\stau$/$\neuto$,
the relic density eventually drops below the WMAP range. Typically
a value of the relic density within the WMAP range requires a mass
difference $\Delta M_{\stau\neuto}\approx 5-15$ GeV for
$\mneuto\approx {\cal O} (100)$ GeV while a smaller mass splitting
is required for heavier neutralinos (near degeneracy is required
for $\mneuto=400$GeV). Fig.~\ref{t1035_coan} shows the relic
density versus the mass difference for $\tan\beta=10,35$ as well
as the mass difference as a function of the neutralino mass. At
$\tan\beta=35$, one can afford a larger mass splitting as there is
a larger contribution from the  s-channel Higgs exchange to
neutralino annihilation into fermions as well as an important
contribution to $\neuto\tilde\tau\ra \tau h$. The latter contains
a term proportionnal to the sfermion mixing.
 The mass difference between the LSP and the sleptons of the first
two generations  can be much larger and ranges from  10-20GeV for
lower values of $\tan\beta$   up to 100 GeV when $\tan\beta\geq
35$. There, the contribution from coannihilation processes with
first generation sleptons become irrelevant {\footnote{Note that
the small mass difference between the stau and the LSP raises the
question of detectability of a nearly degenerate $\stau$ lepton as
well as the feasability of a precise measurement of its mass at a
collider. For recent studies within the context of a linear
collider see \cite{Bambade:dm,Martyn:2004ew}.}}.
 Eventually, as the
neutralino mass increases the coannihilation cross sections also
become too small and one obtains an upper limit on the LSP mass.
When combining this upper limit with the constraint imposed by the
Higgs mass as well as from $\bsgamma$, we find the range of allowed
values for $\tan\beta=10(35)$, $270(342) GeV <\mhf <1050(1250)$GeV
corresponding to neutralinos of $105(138) GeV<\mneuto<448(540)$~GeV.

\begin{figure*}[tbhp]
\begin{center}
\vspace{-1.2cm}
\includegraphics[width=14cm,height=10cm]{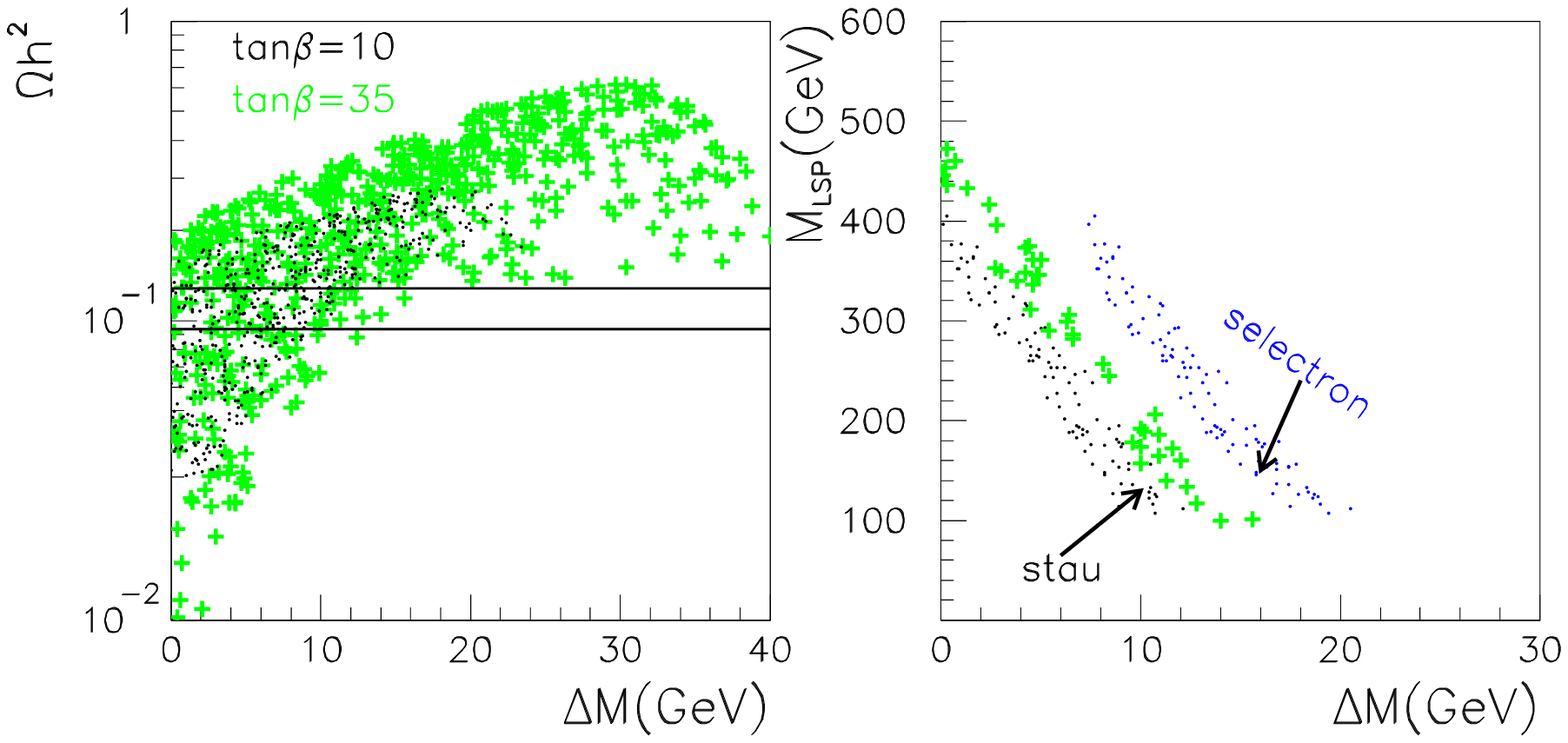}
\vspace{-1.2cm} \caption{\label{t1035_coan}{\em a) $\Omega h^2$ as
a function of the mass difference $\Delta M_{\stau\neuto}$ in the
coannihilation region for
   $\tan\beta=10$(black dots) and $\tan\beta=35$(green crosses).
b) For models where $\Omega h^2$ is within the WMAP range,
$\mneuto$ as a function of the mass difference $\Delta
M_{\stau\neuto}$ for $\tan\beta=10$ (black dots) and
$\tan\beta=35$ (green crosses) and as a function of the mass
difference $\Delta M_{\tilde{e}\neuto}$ for $\tan\beta=10$ (blue
dots). }} \vspace{-.2cm}
\end{center}
\end{figure*}

The Higgs annihilation region includes both a light Higgs
annihilation at low $\mhf$ and a new region which appears only for
large $\tan\beta$. The former occurs only for the LSP mass very
close to  $M_h/2$ since the light Higgs has a very narrow width.
The latter appears  at moderate $\m0-\mhf$ values and corresponds
to the annihilation of neutralinos via the s-channel exchange of a
heavy Higgs. The main channels are $\neuto\neuto \ra
b\bar{b},\tau\bar{\tau}$ corresponding to the preferred decay
channels of the Heavy Higgs. The crucial physical parameter is the
mass difference between $\ma$ and $2\mneuto$ relative to the
pseudoscalar width, hence the importance of including higher-order corrections for the width \cite{Belanger:conf}.  Note that for very heavy scalars the width can
be quite substantial, ${\cal O} (30)$GeV. In
Fig.\ref{higgs_funnel},
  the  mass difference $\Delta M_{A\neuto}=M_A-2\mneuto$ relative
  to the width, $\Gamma_A$, is displayed as function of $\mhf$
 for $\tan\beta=50$.
 Here all constraints are applied.
 Models that fit within the WMAP range
 are roughly within $\Delta M_{A\neuto}<2 \Gamma_A$ except in the low $\mhf$ region.
There  the coannihilation region merges with
the Higgs funnel region. Even though the coannihilation channels never
dominate  for these large values of $\tan\beta$ the contribution of the light
 sleptons to neutralino pair annihilation is sufficient to get an acceptable value for $\Omega h^2$.
  In fact, nearly
degenerate $\stau$/$\neuto$ will lead to a very low value for the
relic density. Due to the heavy Higgs annihilation channel,
neutralinos as heavy as $650$GeV can give reasonable values for
the relic density. This type of model  could be very difficult to
hunt at the LHC as it features a rather heavy  spectrum. Indeed at
the tip of the Higgs funnel, one finds all
 squarks and the gluino in the $2.5-3.2$ TeV range.  Note that  the position of the  Higgs funnel
 is sensitive to the standard model input value used in the RGE codes, in
particular $m_b(m_b)$ \cite{Gomez:mb} as well as
  the top quark mass. Indeed the  soft scalar masses in the Higgs potential,
  $m_{H_1}$($m_{H_2}$),
  from which one evaluates the mass of the pseudoscalar at the EWSB scale
 are very sensitive to the bottom (top)
  Yukawa's \cite{Allanach:codes}.

\begin{figure*}[tbhp]
\begin{center}
\vspace{-1.2cm}
\includegraphics[width=14cm,height=10cm]{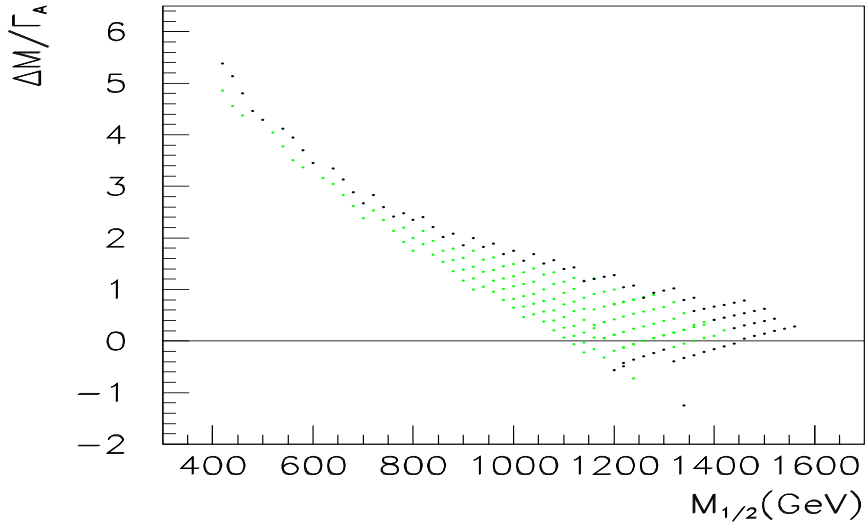}
\vspace{-1.9cm} \caption{\label{higgs_funnel}{\em Mass difference
relative to the pseudoscalar width ($\Delta M_{A\neuto}/\Gamma_A$) as
function of $\mhf$ for $\tan\beta=50$, $\mu>0$, $A_0=0$.
Constraints from LEP as well as $\bsgamma$ are taken into account.
Models within (below) the WMAP range are labelled by black (green/grey) dots.
 }} \vspace{-.2cm}
\end{center}
\end{figure*}

The last allowed region is the focus point region, this region is
found at high values of $\m0$ where the value of $\mu$ drops
rapidly. When $\mu<M_1,M_2$,  the LSP has a significant Higgsino
fraction. Furthermore the NLSP ($\neutt$ and/or $\chargopm$) are
also mostly Higgsino and are not much heavier than the LSP. Thus
the neutralinos/charginos coannihilation channels are favoured.
The position of the focus point region is very sensitive to the
value of the top quark mass that enters the RGE and also differs
for different RGE codes
\cite{Allanach:codes,Allanach:houchesrge,nous_preparation}. For
the case of SOFTSUSY used here, rather low values of the top quark
mass are necessary to reach the region where $\mu$ drops rapidly,
at least for intermediate values of $\tan\beta$. As displayed in
Fig.~\ref{sugra_mt175}, for $m_t=175$~GeV, one finds a
cosmologically allowed region at large $\m0$ only for large
$\tan\beta$. In the focus point region, the main annihilation
channels are $\neuto\neuto\ra W^+W^-,ZZ$. Coannihilation channels
such as $\neuto\chargopm\ra q q'$ can also help reduce the relic
density. One also finds  an important contribution from
$\neuto\neuto\ra t\bar{t}$ as soon as this channel becomes
kinematically accessible. This occurs, in the focus point region,
when $\m0\approx 4.5-5$TeV, Fig.~\ref{sugra_mt175}d. Even though
the stops are heavy ($\msto\approx 2.5$TeV), a sufficiently large
annihilation cross section results from the Z exchange
contribution. For the gauge bosons annihilation channels, values
consistent with WMAP for $\Omega h^2$ require not so large a value
for $\mu$. This implies that the chargino and lightest neutralinos
should be rather light. For example,  the chargino and neutralino
are confined to have $\mchargo< 375$GeV for $\tan\beta=50$,
$\mt=175$~GeV.  The
 chargino and neutralino can be nearly degenerate ${\cal O}(10)$GeV but this
 occurs mainly when the relic density falls below the WMAP range.
  For $\Omega h^2\approx .1$ typical mass differences are rather around
  $50$GeV.
 Note that models in the focus point
region feature a very heavy sfermion sector, the LHC will have
little opportunity to discover the squarks, only the gluino could
be accessible. At the same time the rather light Higgsino sector
could be probed at a linear collider, extending the reach of the
LHC \cite{Baer:lc}. Furthermore these models are also interesting
for future direct dark matter detection experiments as will be
discussed in Section~\ref{direct}.

\begin{figure*}[tbhp]
\begin{center}
\includegraphics[width=16cm,height=10cm]{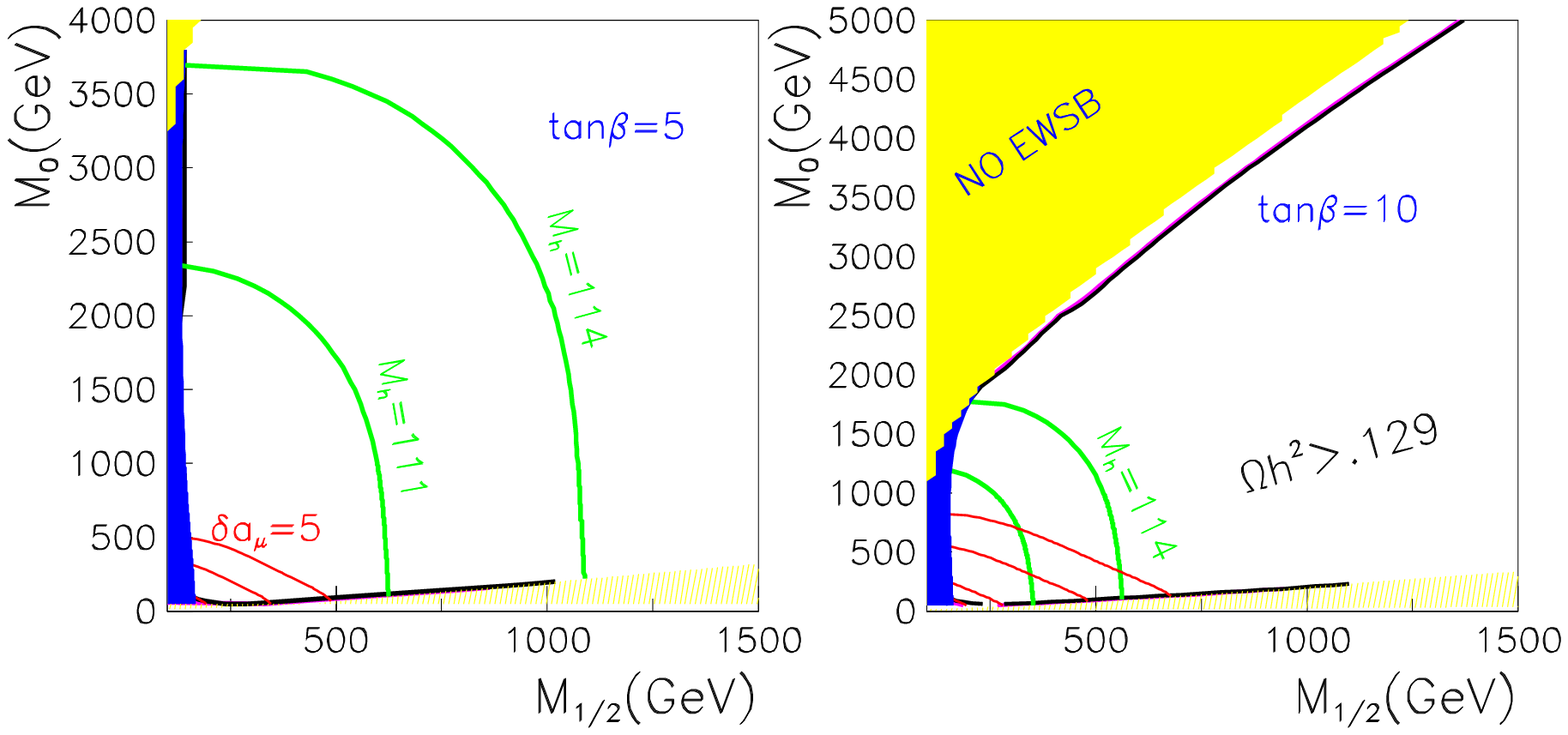}
\includegraphics[width=16cm,height=10cm]{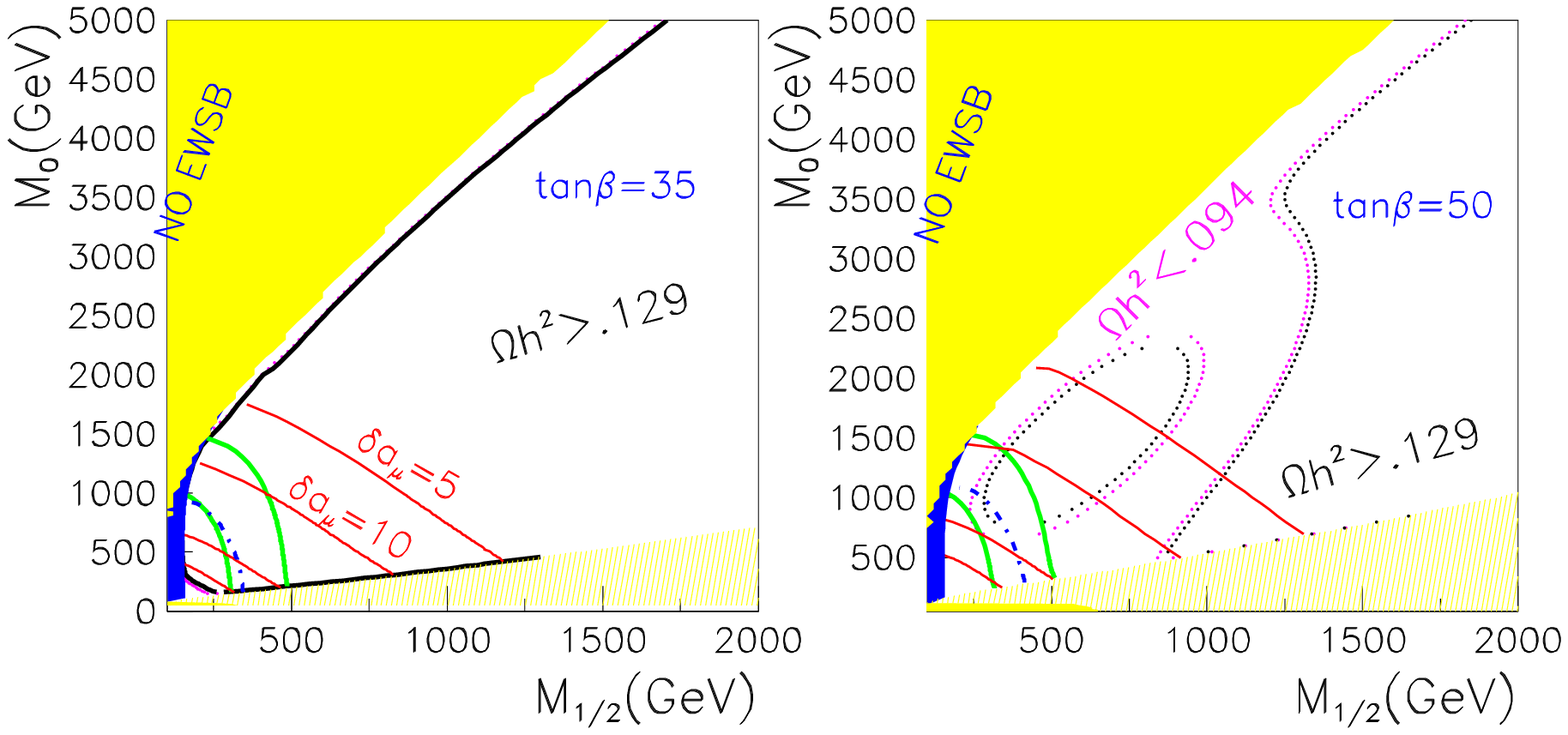}
\vspace{-1.9cm} \caption{\label{mt170}{\em Same as in Fig.~2 for
$\mt=170$~GeV}} \vspace{-.2cm}
\end{center}
\end{figure*}

\subsubsection{$\mt\neq 175$~GeV}

The position of the allowed region at large $\m0$ is the main
qualitative difference between mSUGRA models with a lower input
values for the top quark mass while the more stringent constraint
originating  from  the light Higgs mass shifts the range of
allowed values for sparticle masses  in the low $\m0$ region. For
example, when $m_t=170$GeV, one finds a focus point region even at
$\tan\beta=10$, Fig.~\ref{mt170}. At large values of $\tan\beta$,
the focus point  region  becomes much wider. Indeed all points
between the contour delimiting the region forbidden by the
requirement of electroweak symmetry breaking (EWSB) and $\Omega
h^2=.094$ have a value for the relic density below the WMAP range,
indicating eventually some other dark matter component. The shape
of the allowed region at $\tan\beta=50$ differs significantly from
the case with heavier top quark mass. First, one tends to get a
lighter heavy Higgs, extending the Higgs funnel region towards
higher values of $\m0$, at the same time shifts in $\mu$ are found
pushing the focus point towards lower values of $\m0$. This is
solely an effect of the RGE dependence of the pseudoscalar mass
and of $\mu$ on the top quark mass. As concerns the relic density
constraint the requirement remains the same, only models where the
neutralino lies near $M_A/2$, roughly $\Delta M/\Gamma_A\approx 2$
are allowed.  The Higgs funnel and focus point regions merge for
$\mhf\approx 1$~TeV. Furthermore, because of the additional
contribution from Higgs exchange, one finds acceptable values for
the relic density in the low $\m0-\mhf$ region.
  In the end, a large fraction of parameter space is
allowed by the relic density constraint. Of course direct
measurements such as $\bsgamma$ play an important role here
particularly in the low $\m0-\mhf$  region.

Conversely a heavier top quark, say $m_t=179$GeV, means a shift of
the focus point region towards higher values of  $\m0$
\cite{Baer:2004qq}. We have stopped our scans at $\m0=5$~TeV, as
we found  it  rather difficult to obtain a solution to the RGE
that converged rapidly for very large $\m0$ \footnote {The
predictions of the different codes for the SUSY spectrum
calculation can be very different in this region
\cite{Allanach:houchesrge}. For example, with ISAJET the authors
of \cite{Baer:2004qq} find the focus point region all the way up
to $M_0=10$TeV.}}. The relaxed bound from the Higgs mass also
favours the region of parameter space where not so heavy
neutralinos/charginos and sleptons can be found, here $\mneuto>
90$~GeV for $\tan\beta=10$. Finally the Higgs funnel moves closer
to the coannihilation region, indeed the heavy Higgs masses are
shifted upwards with the increase of the top Yukawa making it
increasingly difficult to have $\mneuto\approx M_A/2$. Then for
$\tan\beta=50$  the allowed region only extends slightly above the
stau coannihilation region.

\begin{figure*}[tbhp]
\begin{center}
\vspace{-1.2cm}
\includegraphics[width=16cm,height=10cm]{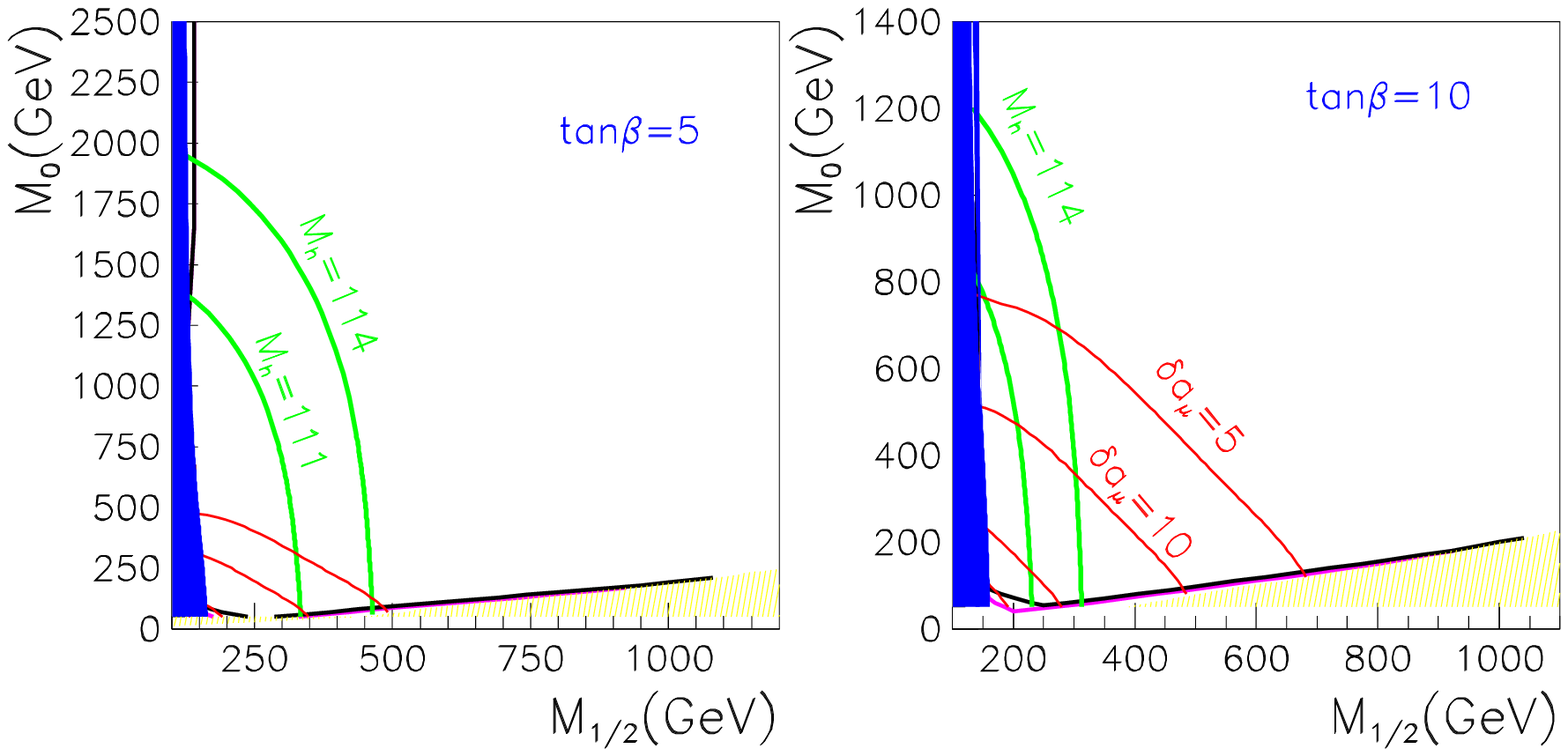}
\includegraphics[width=16cm,height=10cm]{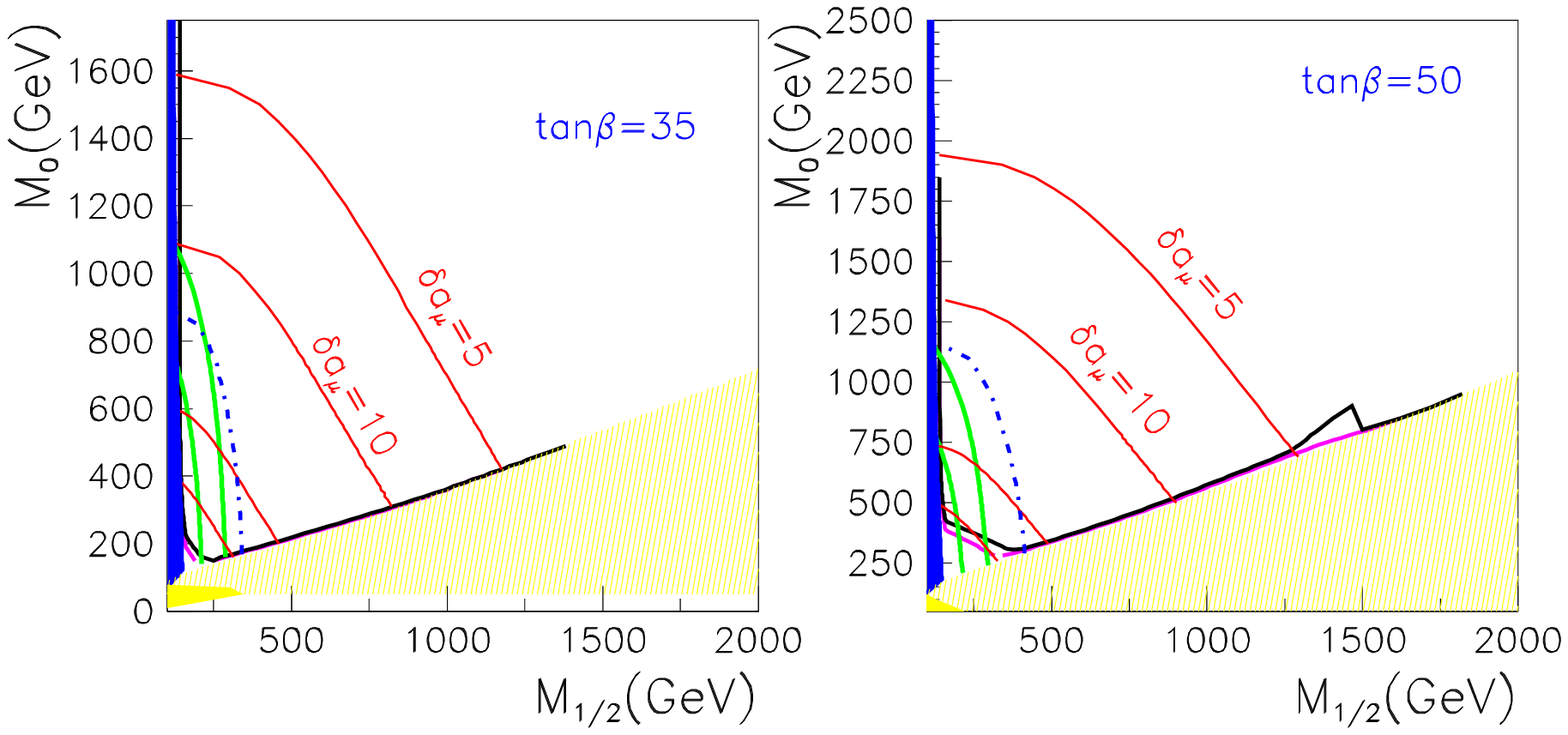}
\vspace{-1.9cm} \caption{\label{mt179}{ \em Same as in Fig.~2 for
$\mt=179$~GeV.
 }} \vspace{-.2cm}
\end{center}
\end{figure*}

\noi {\bf mSUGRA, the case $A_0\neq 0$}

\begin{figure*}[tbhp]
\begin{center}
\vspace{-.2cm}
\includegraphics[width=14cm,height=10cm]{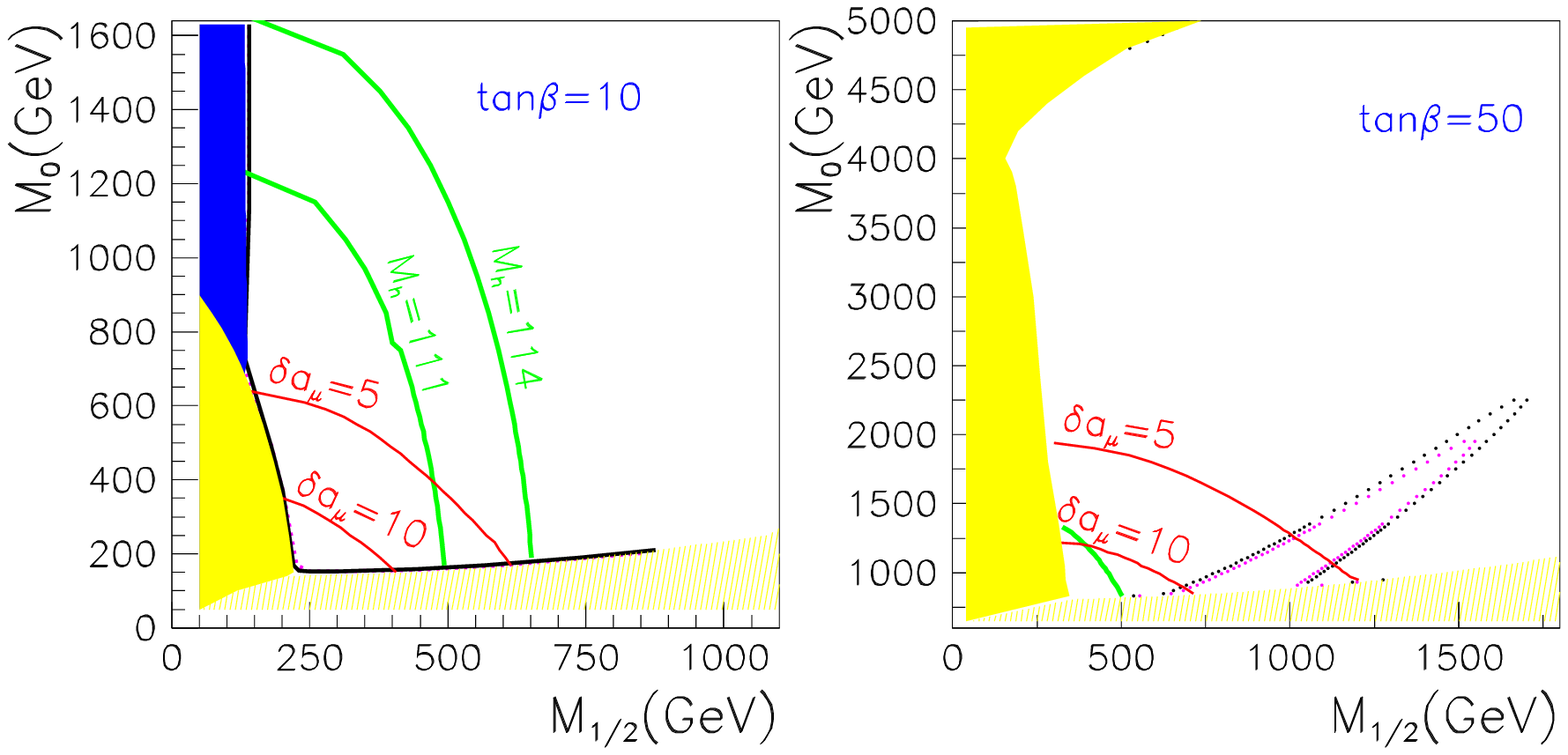}
\vspace{-.9cm} \caption{\label{a2000}{\em Same as in
Fig.~\ref{sugra_mt175} for
 $\mu>0$,  $\mt=175$~GeV and $A_0=2000$GeV. Only the contours
 $\amu=5,10\times 10^{-10}$ (top to bottom) can be seen.
 }}
  \vspace{-.2cm}
\end{center}
\end{figure*}

 For simplicity we have up to now fixed the value of the trilinear coupling to $A_0=0$.
 Although the constraints are not
extremely  sensitive to this parameter
 yet as one varies $A_0$ over the full range $|A_0|<2$TeV, one sees an impact
 on the allowed regions in the $\m0-\mhf$ plane. For one,
 a non-zero value for the trilinear coupling shifts the
physically allowed region where the neutralino is the LSP. First,
a region with  tachyons appears at low $\mhf$. Furthermore, the
stau mass is lowered since the trilinear mixing  affects the
running of the $\tilde\tau_{L,R}$ masses as well as the running of
$\mu$ which determines the mixing in the stau sector. Indeed the
mixing is proportionnal to $A_\tau+\mu\tan\beta$ so is usually
dominated by the term $\mu\tan\beta$. The boundary of the
coannihilation  region then moves towards higher values of $\m0$
for non-zero values of $A_0$. This is particularly obvious at
large $\tan\beta$, Fig.~\ref{a2000},~\ref{a-1000}.

First consider models with $A_0=2000$~GeV and $\mt=175$~GeV. One
finds that consistency with WMAP is limited to a region just at
the boundary of the theoretically excluded region for
$\tan\beta=10$, see Fig.~\ref{a2000}. One finds a new feature in
these scenarios, a region where the $\tilde{t}_1$ is the NLSP.
This region which occurs  at moderate values of $\m0$  lies just
to the right of the region where one finds tachyonic solutions, i.e. for $200 <\m0<650$~GeV.
The main annihilation  for neutralinos is into fermion pairs and
coannihilation channels such as $\neuto \tilde t_1\ra t g,
W^+\bar{b}$    contribute also significantly. Note that the
 contribution from $\neuto \tilde t_1\ra W^+\bar{b}$
 can give the dominant contribution to the
  thermally averaged  cross section exceeding the QCD channels with gluon emission
  {\footnote {This contribution is not included in \cite{abdel_stop}.}}.
 Then one can
satisfy the relic density constraint because of the coannihilation
processes involving stop quarks.  However this stop coannihilation
region is not allowed by  the light Higgs mass constraint. This
constraint is more severe than in models where the trilinear
coupling vanishes, only the stau coannihilation region remains
once the $\mh$ limit is imposed.
 The allowed values for neutralinos masses in the
 coannihilation region
lie in the range $170(126)<\mneuto<440(700)$GeV for
$\tan\beta=10(50)$.
%As usual, NLSP-LSP mass differences ${\cal
%O}(10)$GeV are necessary for low $\tan\beta$.
Of course one can have models with an even heavier LSP since for
 $\tan\beta=50$, both the focus point
region as well as the Higgs funnel are present. These regions are
shifted as compared to the $A_0=0$ case, in particular the heavy
Higgs masses are shifted down making for a more efficient
annihilation of neutralinos via a heavy Higgs exchange.   Then
neutralinos as heavy as $\mneuto\approx 740$~GeV are allowed.

 For negative values of
$A_0$, say $A_0=-1$TeV, the constraints from the Higgs mass are
relaxed while the $\bsgamma$ is more important as was discussed in
the previous section. The relic density is the main constraint on
models with low $\tan\beta$. Again one finds a region where the
$~\tilde t_1$ is the LSP at low $\m0-\mhf$ values. Just when the
$\tilde t_1$ becomes the NLSP, values of the relic density
consistent with WMAP are found, see Fig.\ref{a-1000} for
$\tan\beta=10$. However this region is barely compatible with the
Higgs mass limit and  does not pass the $\bsgamma$ constraint. One
also finds as usual a thin $\stauo$ coannihilation region.
Furthermore, with the large negative mixing which tends to
increase the heavy Higgs masses, the Higgs funnel is reduced to a
tiny region even at $\tan\beta=50$, Fig.~\ref{a-1000}. Note also
that the constraint from $\bsmu$ starts to restrict the mSUGRA
parameter space although it is not yet competitive with other
direct constraints. For larger values of the mixing parameter, e.g.
$A_0=-1700$~GeV, it is possible to find models that satisfy all
constraints and where the $\tilde{t}_1$ is the NLSP
\cite{Santoso:stop}.

\begin{figure*}[tbhp]
\begin{center}
\vspace{-1.2cm}
\includegraphics[width=14cm,height=10cm]{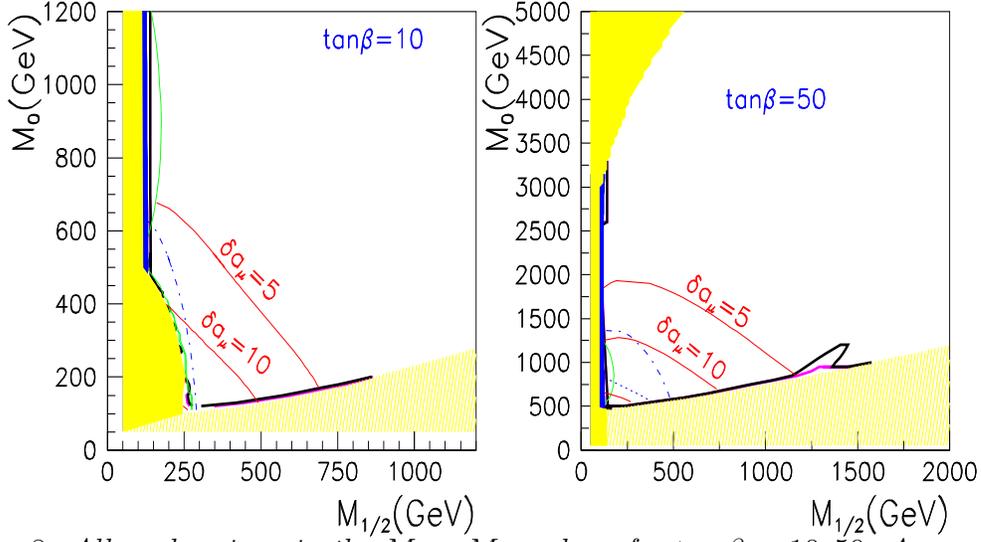}
\vspace{-1.9cm} \caption{\label{a-1000}{\em
 Allowed regions
in the $\m0-\mhf$ plane  for $\tan\beta=10,50$, $A_0=-1000$~GeV,
$\mu>0$ and $\mt=175$~GeV. Contours for  $\Omega h^2= .129 (.094)$
(black/pink dots) $\mh=114$GeV (green) $\amu=5,10,30\times
10^{-10}$ (red) (top to bottom), $\bsgamma=2.25 \times 10^{-4}$
(dash-dot/blue) and $\bsmu=2\times 10^{-7}$ (dots) for
$tan\beta=50$. The exclusion regions are labelled in the same
manner as in Fig.~\ref{sugra_mt175}.
 }} \vspace{-.2cm}
\end{center}
\end{figure*}

\subsection{$\chi^2$ fit}

\begin{figure*}[tbhp]
\begin{center}
\vspace{-1.2cm}
\includegraphics[width=14cm,height=10cm]{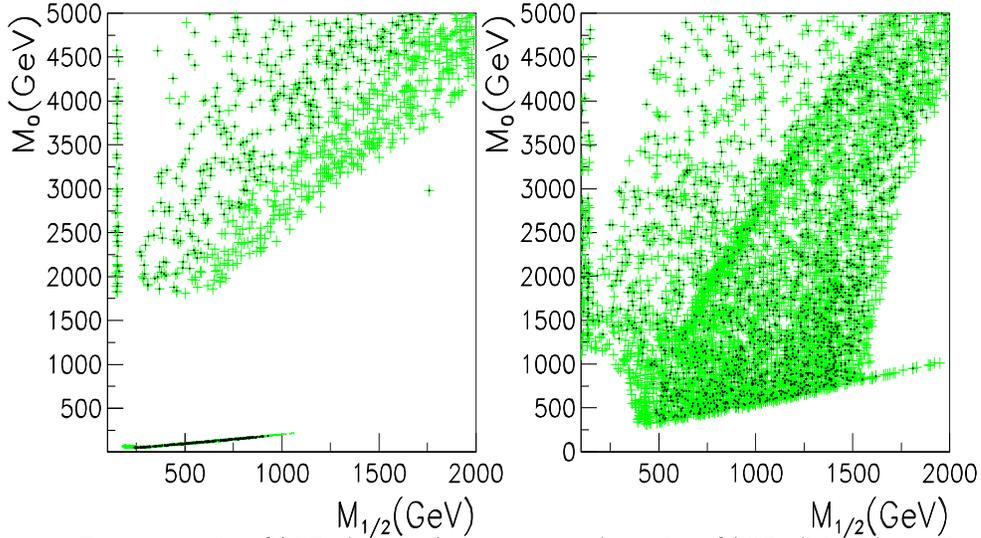}
\vspace{-1.9cm} \caption{\label{chi2}{
Regions of 95\%CL (green/grey crosses) and 68\%CL (black) in the $\m0-\mhf$ plane
 from a $\chi^2$ fit to $\Omega h^2$,
 $\mh$, $Br(\bsgamma)$ and $\mt$ for $\tan\beta=10$ (left)
 and $\tan\beta=50$ (right),
  here $\mu>0$ and $A_0=0$.
 }} \vspace{-.2cm}
\end{center}
\end{figure*}

So far we  have  discussed the effect of each type of constraint
taken individually as well as their sensitivity on the top quark
mass. To get a better estimate of the impact of the uncertainty on
the input parameters, we have also performed a $\chi^2$ fit to the
four observables: $\Omega h^2,\bsgamma$, $m_h$ and $m_t$
 for a fixed value of $\tan\beta$ and $A_0$.
We use  $\mt=174.3\pm 5.1$~GeV.
 We
show in Fig.~\ref{chi2} the allowed area at $68\%$ and $95\%$ C.L.
One finds  that the favoured regions are the coannihilation as
well as the focus point region for $\tan\beta=10$. The area of
parameter space allowed at $95\%$C.L. after allowing for the $\mt$
uncertainty now seems very wide especially at $\m0>2$~TeV. This
reflects the high sensitivity of the RGE to the value of the top
quark mass. For $\tan\beta=50$, the $\stauo$-coannihilation, Higgs funnel
and focus point region are not easily distinguishable and a large
fraction of parameter space is allowed. Note that both the Higgs
funnel and focus point regions are very sensitive to the value of
the top quark mass as we have seen in
Fig.~\ref{sugra_mt175},\ref{mt170},\ref{mt179}.

In summary, in mSUGRA models a combination of the direct collider
limits and relic density constraints, restricts the allowed models
in very specific regions of parameter space. The exact location of
these regions in the $\m0-\mhf$ plane is however sensitive to
theoretical uncertainties in the evaluation of the spectrum.
Nevertheless the predictions at the EW scale are rather contrived:
either nearly degenerate $\stau$/neutralinos,   light
higgsino-type neutralinos or at large $\tan\beta$, neutralinos
that are near $M_A/2$.  The results we have presented agree
qualitatively with other published results, however differences in
the position of the heavy Higgs funnel as well as  the focus point
region are found \cite{Ellis:wmap, Baer:2003wx}. To a large extent
these differences can be traced back to differences in the
evaluation of the supersymmetric spectrum at the EW scale. These
issues were brought up in Ref.~\cite{Allanach:houchesrge} and more
details will appear in \cite{nous_preparation}.

\section{Non-universal gauginos}

In models where one relaxes the gaugino universality conditions
the allowed regions in parameter space  can change drastically
\cite{Baer:nonuni,arnowittnonuni,Munoz:nonuni}. For one, the wino
content of the LSP can be significantly increased, favouring the
annihilation cross section into pairs of gauge bosons. This occurs
for example for $M'_1/M'_2>1$ at the GUT scale.
 Then reasonable values for
the relic density can be obtained even for heavy sleptons. The
annihilation cross section into pairs of W tends to be more
efficient than in lepton pairs \cite{Birkedal-Hansen:2003gy} in
fact so efficient that the relic density often falls below the
WMAP range. Furthermore, the coannihilation with heavier charginos
and neutralinos, also the most efficient ones, are much more
frequent in this type of model. Second, as in the universal case,
one finds significant areas of parameter space where neutralino
annihilation proceeds through a s-channel Higgs exchange. However
these regions do not necessarily only appear at large $\tan\beta$.
Allowing $M'_3/M'_2< 1$ at the GUT scale reduces  the value of the
heavy Higgs masses without much effect on the neutralino mass.
This type of models also predicts smaller values for $\mu$ which
means a LSP with a significant Higgsino component and more
efficient annihilation channels without requiring  high
$\tan\beta$ \cite{Nezri:nonuni}. Alternatively, reducing
$M'_1/M'_2$ allows annihilation of neutralino through a light
Higgs exchange. Finally non-universality in the gaugino sector can
also affect the predictions for the sfermion masses. In particular
reducing the parameter $M'_3$ for given $M'_1,M'_2$ values,
produces much lighter squarks. Coannihilation channels with
squarks then are more likely especially when there is a large
mixing in the squark sector. Altogether one finds the same
mechanisms as in mSUGRA for getting sufficient
annihilation/coannihilation of neutralinos but the
 allowed region in the parameter space of the MSSM shifts significantly
 and no longer requires a fine tuning of parameters.
 The direct constraints are of course also modified in non-universal models. This
will be detailed next and illustrated with a few  typical case
studies. Here we only consider the case $m_t=175$GeV.

\subsection{$M'_3 > M'_2$, the case $M'_3=2 M'_2=2 M'_1$}

\begin{figure*}[tbhp]
\begin{center}
\vspace{-1.2cm}
\includegraphics[width=16cm,height=10cm]{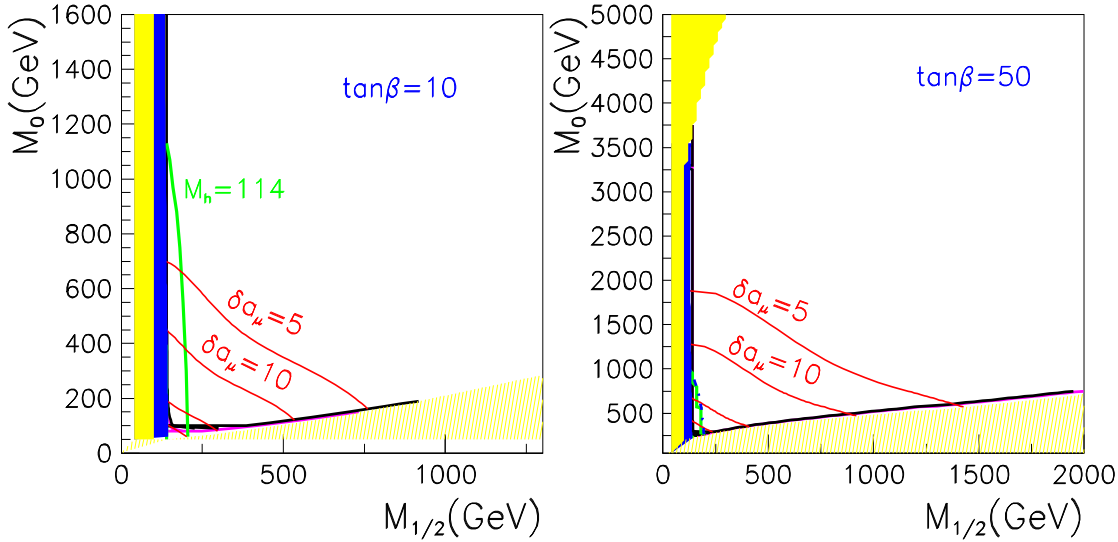}
\vspace{-1.9cm} \caption{\label{m32m2}{\em Same as in
Fig.~\ref{sugra_mt175} for $\mu>0$, $A_0=0$, $\mt=175$ and
$M'_3=2M'_2$. Only contours for $\mh=114$~GeV (green) are
relevant.
 }} \vspace{-.2cm}
\end{center}
\end{figure*}

By changing only $M'_3$ while keeping $M_2'=M_1'$ at the GUT scale
means that at the weak scale one retains the usual  relation  $M_1
\approx 0.4 M_2$. Futhermore the parameter $\mu$ tends to be
larger than in mSUGRA, see Eq.~\ref{mu}. Therefore  the $\neuto$
is nearly a pure bino. The relic density constraint then becomes
more severe than in the universal case since  annihilation of bino
LSP's into light fermions, although dominant, is not sufficient to
bring $\Omega h^2$ in the WMAP range. The pseudoscalar mass, which
is driven by $M_3$, is also much heavier than in the universal
case
  so  annihilation  through the heavy Higgs cannot take place.
  The focus point region is pushed towards higher values of $\m0$ even
  at $\tan\beta=50$.  Then the only allowed region is the $\stauo/\neuto$ coannihilation
  region. This region is always present since  the relation between  $M_0$ and $M_1,M_2$ which to a large extent determines
  $\mneuto,\mstauo$ has not changed.  The allowed narrow band extends beyond
 $\mhf=2$TeV, see Fig.\ref{m32m2}, so that a rather heavy SUSY spectrum is possible.
 In the coannihilation region one finds roughly $\Delta
 M_{\stau\neuto}\approx 5-10$~GeV for $\tan\beta=10$,  comparable to the values found
 for the universal case. For larger values of $\tan\beta$, the increased contribution
 of many channels, in particular those with a light Higgs, $\neuto\stau\ra \tau h,\stau\stau\ra hh$
 means that larger mass differences can be compatible with WMAP.

 On the other hand the constraints from colliders are not as severe.
As the Higgs mass is driven by the squark mass which  increases
with $M'_3$,  lower values of $\mhf=M'_2$ should be allowed. The
upper bound on the Higgs mass is indeed easily satisfied for
$\tan\beta=10$ as can be seen in Fig.\ref{m32m2}. Allowed
regions are found even at $\tan\beta=5$ . There for $\mh>111$~GeV
one recovers a small allowed region with $\mhf>250$ GeV and light
sleptons.
 The constraints from precision measurements  are easily satisfied
 in this model since
heavy squarks also imply that $\bsgamma$ is close to the standard
model prediction. Only a tiny region is excluded for
$\tan\beta=50$. In this type of model, for example when
$\mhf>800$~GeV, the heavy squarks make it difficult for the LHC to
discover supersymmetry. On the other hand one recovers part of the
low $\mhf$ region that is  favourable for collider searches of
charginos and sfermions.

\subsection {$M'_3 < M'_2$, the case $M'_3=M'_2/2$}

\begin{figure*}[tbhp]
\begin{center}
\vspace{-1.2cm}
\includegraphics[width=14cm,height=10cm]{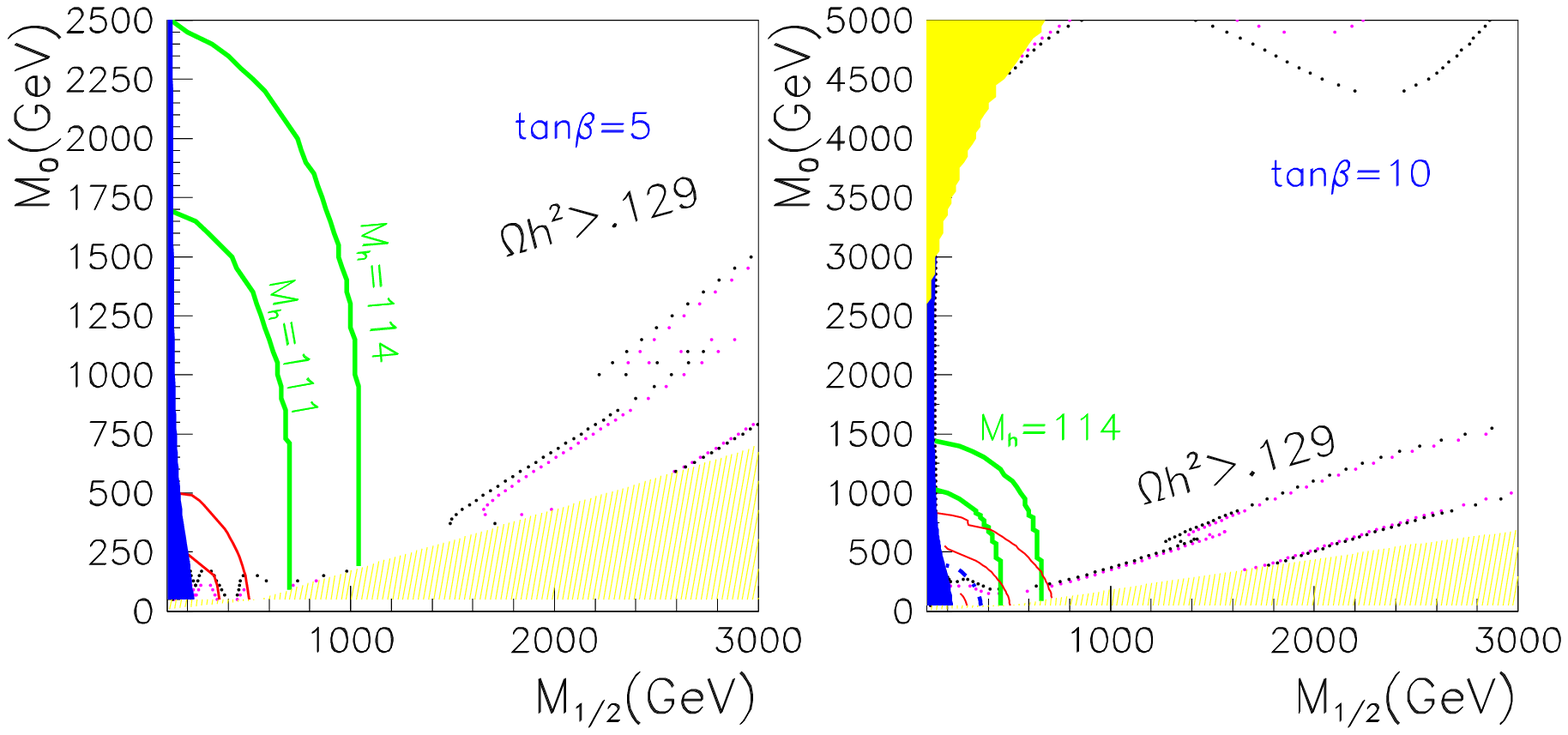}
\includegraphics[width=14cm,height=10cm]{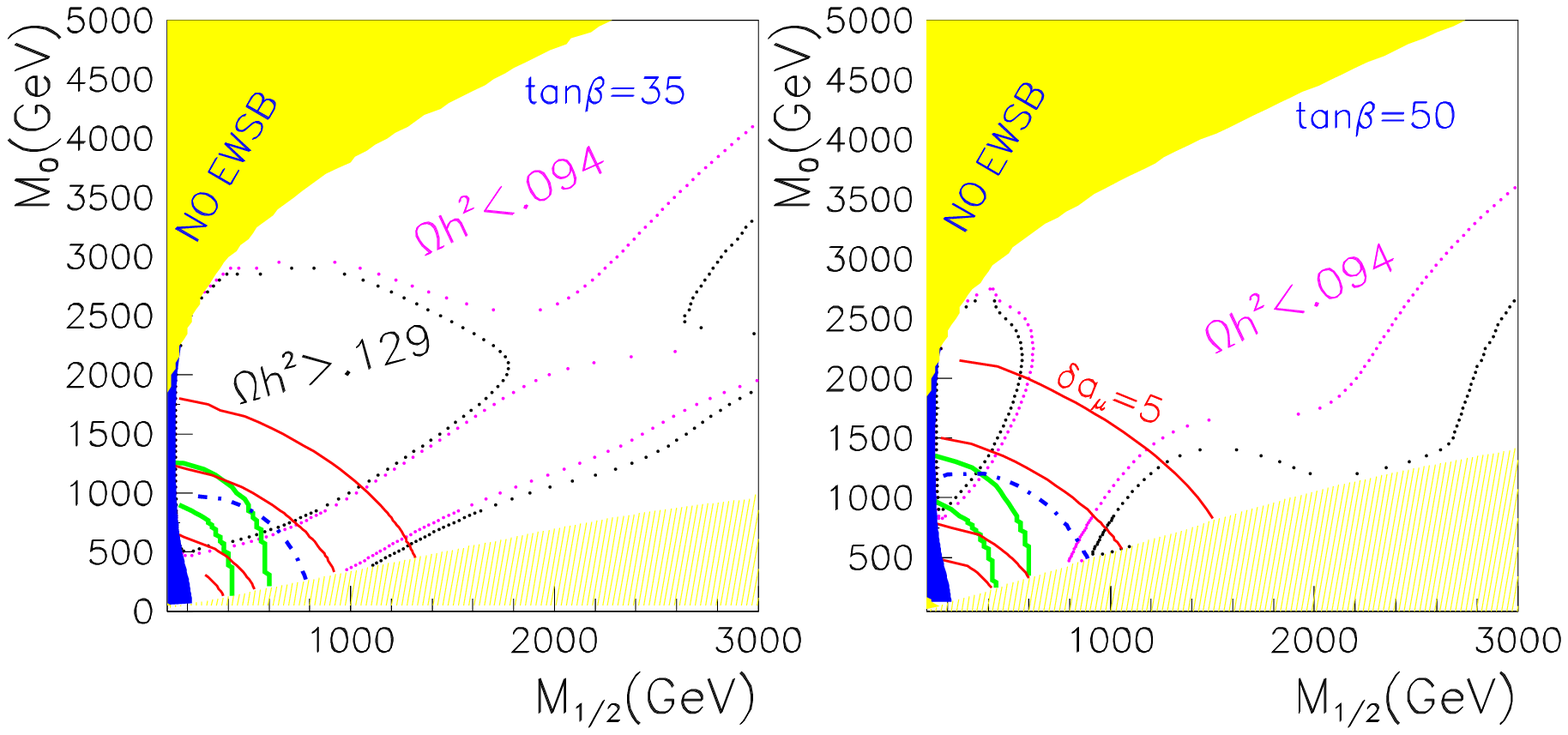}
\vspace{-1.cm} \caption{\label{m3m22}{\em Same as Figure 2
 for $\mu>0$, $A_0=0$, $\mt=175$~GeV and $M'_3=M'_2/2$.
 }} \vspace{-.2cm}
\end{center}
\end{figure*}

Decreasing the value of $M'_3$ while keeping $M'_2=M'_1$ at the
GUT scale means that the relic density constraint  can be
satisfied more easily. First, one finds a neutralino-LSP  with
higher Higgsino component, although the LSP remains mainly  a
bino. Indeed $\mu$ is typically smaller than in unified models,
Eq.\ref{mu}. More importantly one feels the presence of the heavy
Higgs pole even for moderate values of $\tan\beta$. As we have
discussed in Section~\ref{model}, solutions for $M_A \approx 2
\mneuto$ can be found even for $\tan\beta=10$, Eq.~\ref{ma}.
 Neutralino annihilation into fermions is then
dominated by the s-channel scalar exchange and the main channels
are $\neuto\neuto\ra t\bar{t},b\bar{b},\tau\bar\tau$, the former
being dominant at lower values of $\tan\beta$ only. Note that for
$\tan\beta=10$ the coupling of the pseudoscalar to $b$ quarks is
already  2 to 3 times larger than to t quarks the ratio $\frac
{Abb}{Att}\approx\frac{m_b}{m_t}\tan\beta^2$ and this mode starts
to dominate. The region that in the universal case was called the
$\stau$ coannihilation region now becomes a Higgs/coannihilation
one and
 includes the low $\m0-\mhf$ region, similar to what was observed
previously for large values of $\tan\beta$.
 In fact as one moves closer to the $\tilde\tau$ LSP region
  and coannihilation
 channels start to contribute significantly to the effective cross section, the relic density
 is already much below the WMAP range. This corresponds in
 Fig.\ref{m3m22} to the region between the $\stau$ LSP line and
 the contour $\Omega h^2=.094$.
This means that values of $\Omega h^2$ within the range of WMAP
are no longer associated with nearly degenerate $\stau-\neuto$.
As in mSUGRA, as one moves towards larger values of $\mhf$
 the stau coannihilation cross sections  become much  smaller and even  models with
 completely degenerate stau/neutralino cannot meet
  the upper limit  of WMAP on the relic density.
However  a new Higgs funnel region appears, it corresponds again
to s-channel Higgs resonance but this time for the coannihilation
processes $\neuto\neutt,\neuto\neutth\ra
t\bar{t},b\bar{b},\tau\bar{\tau}$ as well as charged Higgs
exchange in $\neuto\chargop\ra t\bar{b}$. The onset of the charged
Higgs exchange contribution is easily visible as a kink in the
contours of constant relic density in Fig. \ref{m3m22}a-b. The
focus point region is also more important even for $\tan\beta=10$.

As one increases $\tan\beta$, the heavy Higgs decay into b-quarks
 completely dominates. The allowed region in the $\m0-\mhf$
plane correspond mainly to annihilation into $b\bar{b},
\tau\bar\tau$.  One does not find a region where
$\stau$-coannihilation dominates, although $\tilde\tau-\neuto$ can be
quite close in mass, this occurs only when one is near the
Higgs resonance. Then the relic density is much below the WMAP
range.  For values of $\mhf>1100$GeV, only annihilation or
 coannihilation with heavier neutralinos/charginos into fermion pairs
 through a Higgs exchange take place.
Since the value of $\mu$ is smaller than in the universal case,
there is also a more important  contribution from the annihilation
into gauge bosons channel and the focus point region is much wider
for a fixed top quark mass. Eventually the focus point and Higgs
annihilation region merge, for example for $\tan\beta=35$ and for
$\mhf\approx 2$TeV the entire range of $\m0$ is below the WMAP
upper bound. For $\tan\beta=50$, the merging of the two regions is
complete and a large fraction of the parameter space is allowed.
As for other values of $\tan\beta$, annihilation into $b\bar{b}$
pairs dominates for low values of $\m0$ and annihilation (or
coannihilation) into gauge boson pairs eventually takes over as
one moves close to the region where EWSB cannot take place. We remark that even with
a moderate trilinear mixing, say $A_0=-600$GeV, one finds regions
where the $\tilde{t_1}$ is the NLSP.

As one would expect, when $M'_3 < M'_2$, the Higgs mass constraint
becomes more severe as the squark masses are lower than in the
universal case for a fixed value of $\mhf=M'_2$. For example, for
$\tan\beta=10$, only $\mhf>460GeV$ is allowed in the
coannihilation region. The lighter squarks relative to the
universal case  imply a larger deviation from the standard model
prediction for $\bsgamma$ as well. The exclusion region in the
left-hand corner of the $\m0-\mhf$ plane where the $\bsgamma$
branching ratio drops below the allowed range becomes more
important. As usual this constraint takes over the one coming from
the Higgs mass as one increases $\tan\beta$. For  $\delta a_\mu$,
the main difference from the universal case is that the value of
$\mu$  are modified by the RGE's. Nevertheless the allowed region
is very similar to the one obtained in the universal case, the
contours are only slightly shifted
 towards higher $\mhf$ values. Note that here again requiring a strict deviation from the
 standard model, would rule out most of the region at large $\m0-\mhf$. Combining this with
 other direct constraints would leave only a small region  of parameter space for $\tan\beta\leq 10$
 as the direct constraints are more effective in this class of model.

In these models typically  squarks are lighter than in mSUGRA,  one then expects a
good discovery potential for the LHC. However
 as one easily finds a region where $\mneuto\approx \ma/2$, even
for very heavy neutralinos one can find reasonable values for the
relic density. It is possible then to obtain a very heavy
supersymmetric spectrum. For example for $\m0= 4TeV, \mhf=3$TeV,
the spectrum consists of heavy squarks, $m_{\tilde q}=3-4$TeV,
neutralinos and charginos, $\mneuto\approx\mchargo\approx 820$GeV
and also heavy Higgses $m_H\approx m_{H^+}\approx m_A\approx
1.5$TeV, a difficult task for discovering supersymmetry at
colliders. The only new particle that could be reached at a
collider would be the light Higgs.
 In these models even  $\amu$ does not differ significantly from zero.
 Only an unambiguous evidence for a non-zero contribution to $\amu$
 would restrict this class of models with heavy sparticles.

\subsection{ The general case $M'_3\neq M'_2=M'_1$}

\begin{figure*}[tbhp]
\begin{center}
\vspace{.2cm}
\includegraphics[width=14cm,height=10cm]{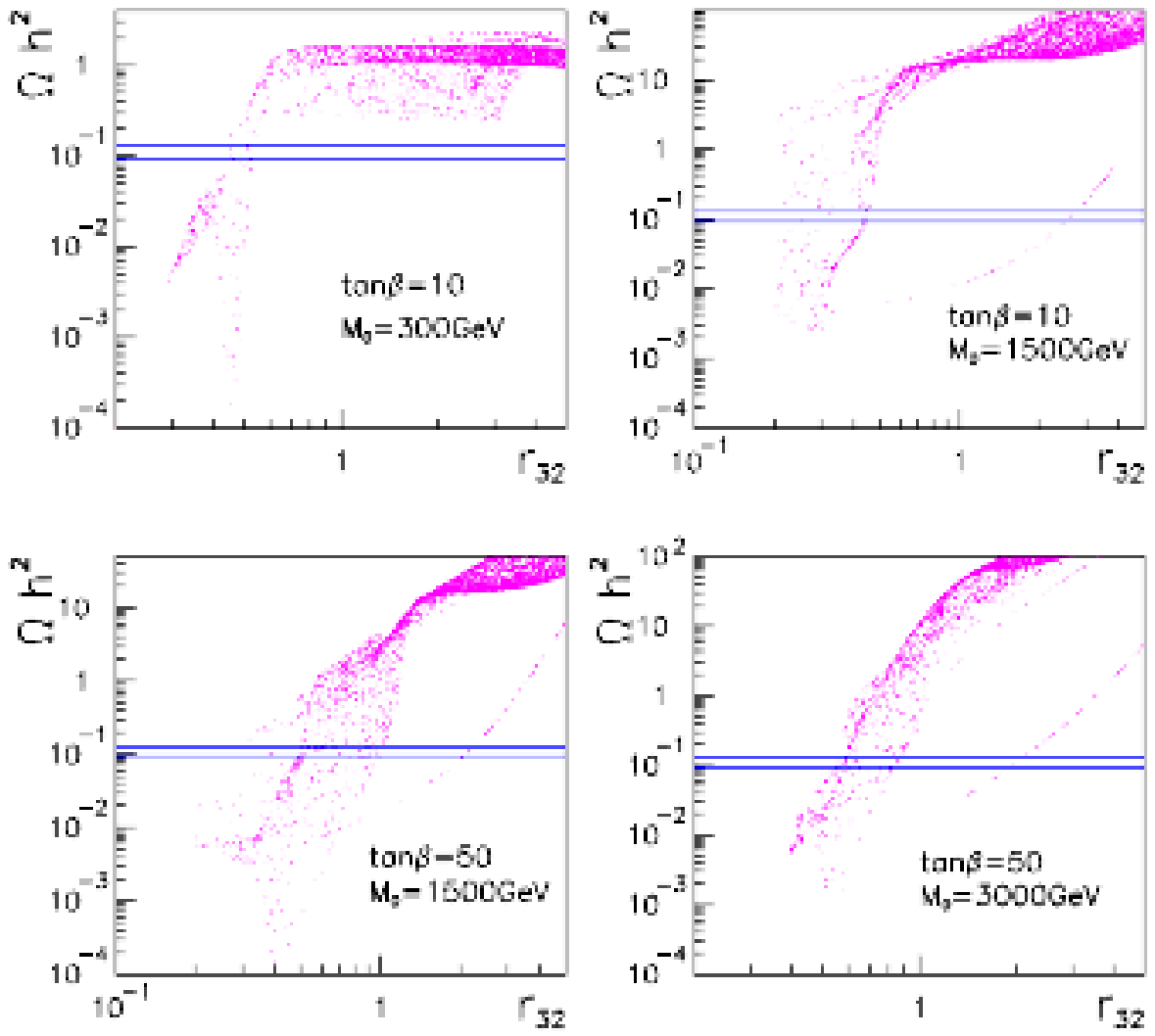}
\vspace{-.5cm} \caption{\label{m3xm2}{\em $\Omega h^2$ vs
$r_{32}$ for $\tan\beta=10$, $\m0=300,1500$GeV and
$\tan\beta=50$, $\m0=1500,3000$GeV. In all cases,
   $\mu>0$,
$A_0=0$, $mt=175$~GeV and the constraints from $\bsgamma$, $\mh$ as
well as LEP direct limits are imposed.
 }} \vspace{-.2cm}
\end{center}
\end{figure*}

There is a smooth transition from the region with universality
where $\Omega h^2$ is in general too large to the region
$M'_3<M'_2$ where there is not enough relic density. This is
illustrated in Fig.~\ref{m3xm2} showing the value of $\Omega h^2$
as function of $r_{32}$ for three typical values of
$\m0=300,1500,3000$~GeV. For these values of $\m0$ one  crosses
the coannihilation, Higgs exchange and focus point region
respectively for $\mt=175$~GeV. Clearly it is much easier to
satisfy the relic density upper bound in models where $M'_3<M'_2$.
As just discussed this is the effect of the heavy Higgs pole and
of the more important Higgsino component of the bino LSP. The
Higgs funnel can be found even for $\m0=300$~GeV as long as
$r_{32}<.5$. Furthermore for $\tan\beta=50$, and $\m0=3000$~GeV,
nearly all models are below the WMAP range when $r_{32}<.4$, this
is a combined effect of a smaller value for $\mu$ and $\ma$ as
compared to the universal case.

% In the universal model the points that pass the
%constraints are those with small sfermion masses (slepton%
%coannihilation) while for lower $M'_3$ all constraint can be
%satisfied   even for large $M_0$.

In conclusion, relaxing the unification condition on $M'_3/M'_2$
while keeping $M'_1=M'_2$ helps recover some values for low
$\tan\beta$ because of either a weaker Higgs mass limit or
possible annihilation through a heavy Higgs resonance. This was
illustrated for $\tan\beta=5$.
 Furthermore, the case $M'_3<M'_2$ satisfies more
easily the relic density constraints for any values of
$\tan\beta$.  In particular for large $\tan\beta$ where the relic
density constraint is satisfied in a large portion of the
available parameter space. A significant fraction  of the
$r_{32}<1$ models then predict squarks within the range accessible
by the LHC. On the other hand because of
annihilation/coannihilation through the heavy Higgses, scenarios
with heavy neutralinos as well as heavy sfermions can satisfy all
constraints while predicting few signals at colliders. In  these
models a value of $\amu$ compatible with the standard model is
predicted, evidence of a deviation might be the best way to rule
them out.

\subsection{The case $M'_1\neq M'_2=M'_3$}

\begin{figure*}[tbhp]
\begin{center}
\vspace{.2cm}
\includegraphics[width=14cm,height=10cm]{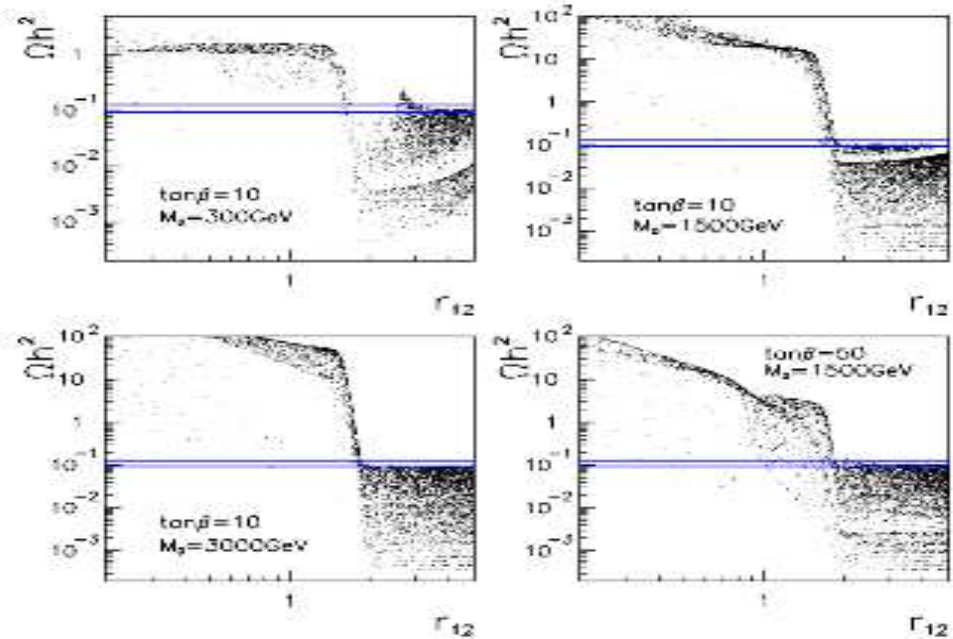}
\vspace{-.5cm} \caption{\label{m1xm2}{\em
 Relic density vs $r_{12}$ in models   with $\mu>0$,
   $A_0=0$, $mt=175$~GeV for a-c) $\tan\beta=10$ and $\m0=300,1500,3000$GeV
d)  $\tan\beta=50$, $\m0=1500$GeV. The horizontal lines denote the
2$\sigma$ region of WMAP.
 }}
  \vspace{-.2cm}
\end{center}
\end{figure*}

We now restore the unification condition $M'_3=M'_2$ while
allowing $M'_1$ to differ from $M'_2$ at the GUT scale. This will
directly affect the wino/bino nature of the LSP. If the LSP has a
higher wino component, which  occurs in  models where $M_1\geq
M_2$ at the weak scale,  annihilation into W pairs becomes
important especially since LEP constraints forces $\mneuto>M_W$.
Furthermore this process is much more efficient than the
annihilation into fermion pairs that dominated for a bino LSP. A
wino LSP also opens up the coannihilation channels with other
gauginos. Then one expects the relic density constraint to be
easily  satisfied in models  where $M'_1>M'_2$. For fixed values
of $\m0$, we scanned over models with $.2<r_{12}<5$ and
$\tan\beta=10,50$. Fig.\ref{m1xm2} clearly shows the dramatic
effect on $\Omega h^2$ of increasing $r_{12}$, varying from almost
always too large relic density when $r_{12}<1$ to nearly always
too small when $r_{12}>>1$ passing by an intermediate region where
$\Omega h^2$ lies within the desired range. This value preferred
by WMAP is around $\r12=1.8$ for $\tan\beta=10$ and corresponds to
$M_1\approx M_2$ at the electroweak scale. Thus the lightest
neutralino is a mixed bino/wino. We will study this case in more
details below. Note that we have stopped our scans at $M_2=2$TeV
so that we have not allowed very heavy neutralinos. Had we
extended the parameter space it would have been possible to find
models with very heavy neutralinos that  fell within the range of
WMAP even for $r_{12}>2$. For higher values of $\tan\beta$ roughly
the same behaviour is observed, only then a wider selection of
models passes the constraint. For example, for $\tan\beta=50$ and
$\m0=1500$GeV, many models satisfy the relic density constraint,
this reflects the presence of the Higgs funnel which appears
already  for $r_{12}\approx 1$.

\subsection {$M'_1 > M'_2$, the case $M'_1=1.8M'_2=1.8M'_3$}

\begin{figure*}[tbhp]
\begin{center}
\vspace{-1.2cm}
\includegraphics[width=14cm,height=10cm]{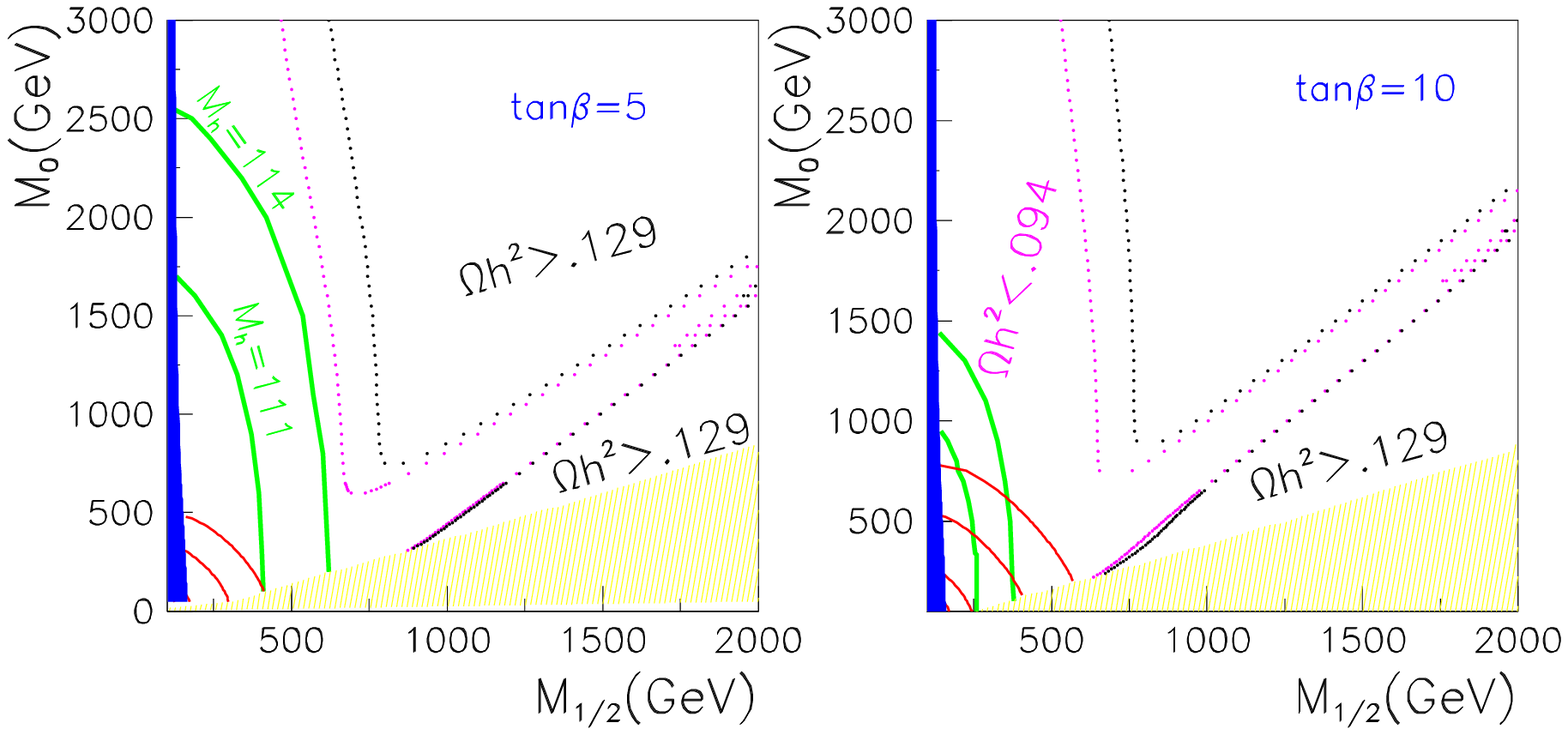}
\includegraphics[width=14cm,height=10cm]{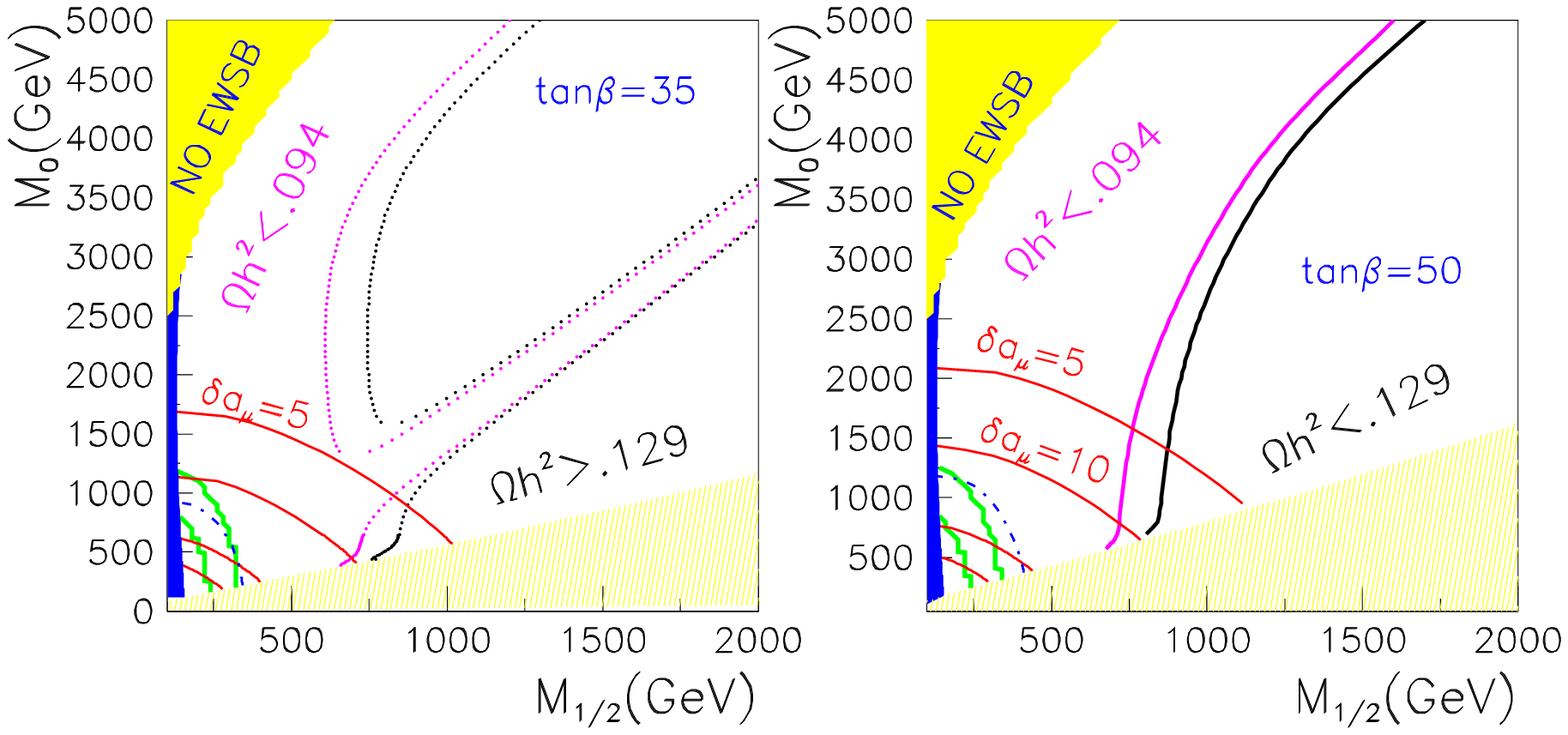}
\vspace{-1.2cm} \caption{\label{m118m2t510}{\em
Same as Figure 10
 for $\mu>0$ ,  $A_0=0$, $mt=175$~GeV, $M'_1=1.8 M'_2=1.8M'_3$.
 }} \vspace{-.2cm}
\end{center}
\end{figure*}

We first consider the special case $M'_1=1.8M'_2$.
 This type of model can arise from string
theory. For example  in models where gaugino masses are loop
suppressed and non-universal one loop contribution become
dominant, models have been constructed where at the one-loop level
the mass $M'_1$ exceeds $M'_2$ and $M'_3$
\cite{Gaillard:1999,Bagger:1999}. We recall that at the
electroweak scale this model corresponds to $M_1\approx M_2$. The
lightest neutralino while still dominantly gaugino  has a large
wino component and is typically heavier than in  mSUGRA. Thus the
increased importance of the neutralino annnihilation into
$W^+W^-$. Furthermore the LSP can be  nearly degenerate with the
chargino and $\neutt$. This favours the gaugino coannihilation
channels. The main channels are into quarks and gauge bosons,
$\neuto \neuto,\neuto\neutt, \chargop\chargom \ra W^+W^-$ and
$\neuto\chargop \ra q\bar{q}, ZW$.
%As compared to the mSUGRA
%model, the LSP while still dominantly gaugino has a higher wino content.
The allowed region is not anymore restricted to the low $M_0$
region as even when squarks and sleptons are very heavy so that
the fermion annihilation channels becomes small the annihilation
into gauge bosons still occurs and has a large enough rate. In
fact this mode is so efficient that this type of model often leads
to a value for the relic density that is too low. For example for
$\tan\beta=10$, in Fig.~\ref{m118m2t510}, the whole region to the
left of the $\Omega h^2=.094$ contour has a relic density below
the WMAP range. Agreement with WMAP then implies a rather heavy
LSP, for example $\mneuto\approx 600$ GeV for $\tan\beta=10$.

For neutralinos up to
roughly $\mhf=700$~GeV, the coannihilation with the $\tilde{\tau_1}$  implies a
relic density below the WMAP range. Above this value  one does not
find an important coannihilation region.   The region between the
stau LSP border and the $\Omega h^2=.129$ contour features too high a value for the  relic density. There the NLSP is a slepton but slepton
coannihilation exchange is not efficient enough for very heavy LSP
as we have seen in the mSUGRA case. Since, for a given $\mhf$ the
LSP is much heavier than in mSUGRA, the scale is sligthly
misleading. As one increases $\m0$ and  the mass of the chargino
becomes more degenerate with the mass of the neutralino then the
coannihilation with chargino becomes significant and there is a
region that satisfies WMAP, in this region one is close to the
heavy Higgs pole. The kink at large $\mhf$ occurs when the
coannihilation process $\neuto\chargop\ra t\bar{b}$ lies near the charged Higgs
resonance. For large $\tan\beta$, the heavy Higgs pole moves
towards lower $\mhf$ and merges with the bulk region as well as
the focus point region.

The constraints on this type of models  from the Higgs mass as
well as from precision measurements are not significantly
different than in the universal model. For example,  the allowed
region starts roughly from $\mhf>300(450)$GeV for
$\tan\beta=10(50)$. The first bound is set from the Higgs mass,
the second from $\bsgamma$. Note that the Higgs mass and the
$\bsgamma$ constrain only models where the relic density  is below
WMAP.

For collider searches,  the spectrum of sparticles as concerns the
coloured sector is not dramatically different then in mSUGRA.
However since the WMAP allowed regions are completely different
and in particular include a Higgs funnel at small $\tan\beta$,
models with very heavy   squarks are perfectly acceptable. For
example, for $\m0=\mhf=2$~TeV and $\tan\beta=10$, all squarks are
in the 3-4~TeV range. These are beyond the reach of LHC.
Furthermore, the LSP tends to be much  heavier than in mSUGRA, more
importantly, there is a small  mass difference between the
chargino and the neutralino which can effect searches.

Had we increased further the ratio $r_{12}$ we would have found a
LSP with an even larger wino component. This would increase the
values for the LSP masses compatible with WMAP. In this way one
can easily reconcile the relic density constraint even with a very
heavy SUSY spectrum. For example if $r_{12}=2.5$ the LSP as well
as the whole supersymmetric spectrum lies above $1.6$~TeV.

\subsection{$M'_1<M'_2=M'_3$}

In that case, the neutralino can be very light as the constraint
on the chargino implies a lighter neutralino than in mSUGRA. A
general analysis of MSSM models with light neutralinos and
including all constraints has shown that neutralinos as light as 6
GeV could be allowed  provided the heavy Higgses were rather light
\cite{Belanger:lowneu,Bottino:lowneu}. In SUGRA type models with
non-universal gaugino masses however the heavy Higgs mass is not
an arbitrary parameter but is deduced from the RGE. Typically one
finds a rather heavy scalar, so one cannot appeal to s-channel
annihilation through a heavy Higgs resonance to bring the relic
density in the desired range, mainly  the light Higgs exchange is
available.  As an example take $M'_1/M'_2=1/3$, the main
annihilation channel  is into a pair of fermions. One finds, in
addition to the $\tilde{\tau}$ coannihilation region, a region
allowed by the relic density constraint, that corresponds to
values of $\mhf$ where the neutralino can annihilate through a
light Higgs pole.
 This region is compatible with the light Higgs
mass limit which is more or less the same as in universal models.
For large $\tan\beta$, as in universal models, annihilation into fermions receives a contribution from the heavy Higgs exchange but this is not sufficient to  allow very light neutralinos.
The  relic density constrains neutralinos to be heavier than $\mneuto \approx 30$GeV.
This type of model offers
good prospect for direct detection of the LSP as well as for
supersymmetry discovery both at the  LHC and at the linear
collider as was discussed in \cite{Belanger:lowneu,Bottino:lowneu}.

\subsection{Other non-universal cases}

Finally there is another class of models that feature
non-universal gaugino masses, models  with $M'_1=M'_3\neq M'_2$
\cite{Nezri:nonuni,Munoz:nonuni,Birkedal-Hansen:2002am}. We do not
discuss these models here as the main impact of relaxing
universality has already been illustrated in the cases studied
above in details. First we  expect that when $M'_1=M'_3>M'_2$ the
relic density would be easily satisfied because of the large wino
component.  Of course, as compared to the case where $M'_3=M'_2$
discussed above, shifts in the value of $m_A$ and $\mu$ will
result from an increase  in $M'_3$. Second we expect models with
$M'_1=M'_3<M'_2$,  to be rather similar to the  $M'_3<M'_2$ models
since   the crucial parameters for the evaluation of the relic
density, such as $\mu$ and $\ma$ are determined by $M'_3$ via the
renormalization group equations.

\section{Prospects for direct detection }
\label{direct}

We have just shown that cosmological measurements provide strong
constraints on many supersymmetric models. Here we discuss the
potential of direct detection searches for the LSP  both in the
mSUGRA model and non-universal models.

Several  experiments  have placed limits on the cross section for
the spin independent neutralino proton scattering, typically  at
the level of $10^{-6}$pb \cite{edelweiss}. There is also one
report of a signal by \dama \cite{dama}. In the near future
results form several upgraded detectors are expected. In fact,
 the first results from  CDMS2 have been published recently \cite{cdms2}
showing already significant improvement on the upper limit.
Finally ton-scale detectors are also planned, such as \zepliniv
\cite{zeplin}, \genius \cite{genius} and \xenon \cite{xenon}.
These large scale  detector will reach a sensitivity between
$10^{-9}-10^{-10}$pb for neutralinos around 100GeV. Typically
detectors have a reduced sensitivity as one increases the
neutralino mass. In parallel, detectors that are also sensitive to
the spin dependent part of the neutralino proton cross section
have obtained their first results, \naiad \cite{naiad}, \simple
\cite{simple} and \picasso \cite{picasso}. At the moment  these
detectors are  much less precise and typically limits at the level
of a few pb's are given. Upgrades of these detectors are planned
and eventually a ton-size detector such as \picasso~ could reach a
sensitivity between $10^{-5}-10^{-4}$~pb.

We follow the usual practice and use  the cross section for
annihilation of neutralino on
 proton, $\sigma_{{\neuto}p}$, to compare the direct detection  potential in a given model.
 We do not rescale the cross section to take into account the possibility that the local density
  of neutralino could be lower than expected. However one has to keep this point in mind
  especially for the models where the relic density is very low.
The spin-independent cross section is dominated by the Higgs
exchange and squark exchange diagrams
 whereas the spin dependent one
 receives contribution from the Z exchange and squark exchange diagrams.
One therefore expects to find the largest cross sections, for both
spin-dependent and spin-independent processes,  in the low $\m0$
region where light squarks are found as well as in the focus point
region of mSUGRA models \cite{Baer:dd} where the coupling of neutralinos to both
the light Higgs and the  Z are enhanced because of the mixed
bino/Higgsino nature of the neutralino. Note that in the focus
point region,
 the large coupling to the Z/Higgs implies  a large  annihilation
cross sections of  neutralinos as well as large neutralino proton
cross sections.

\subsection{mSUGRA}
\begin{figure*}[tbhp]
\begin{center}
\vspace{-.2cm}
\includegraphics[width=14cm,height=10cm]{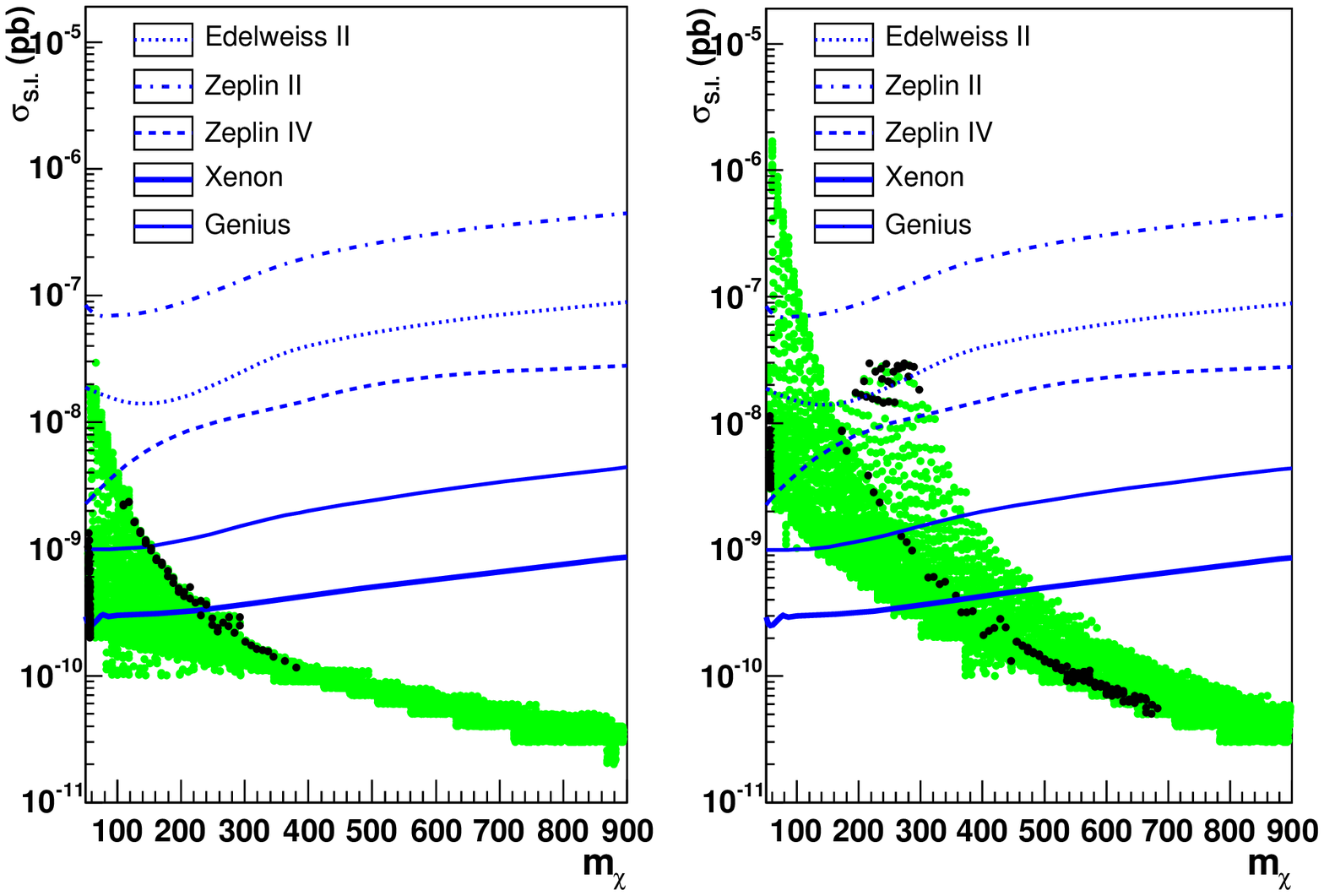}
\includegraphics[width=14cm,height=10cm]{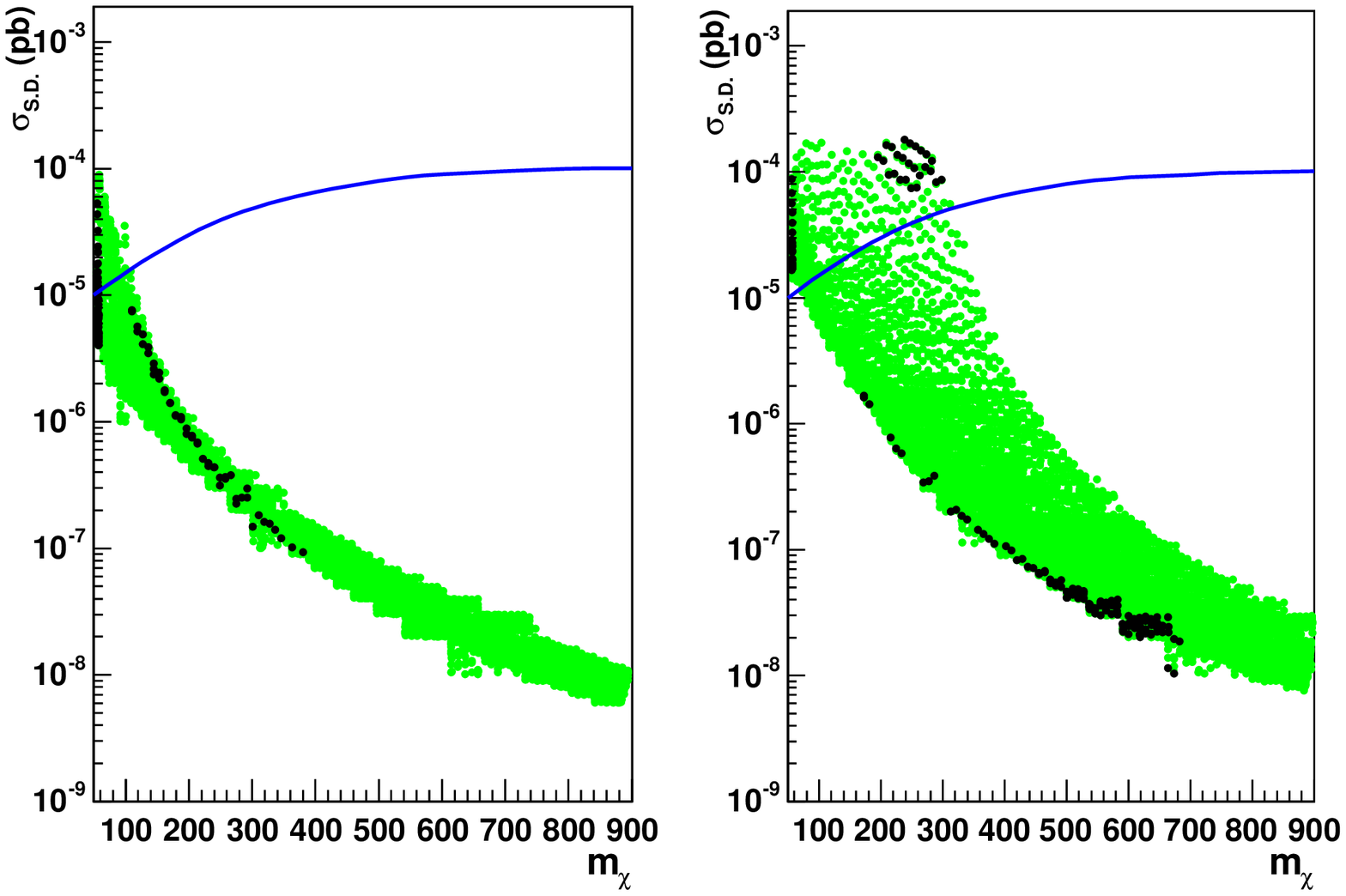}
\vspace{-.1cm} \caption{\label{dd_sugra}{\em Spin independent
(top) and spin dependent (bottom) cross sections for scattering of
neutralinos on proton as fonction of $\mneuto$ for $m_t=175$GeV,
$\tan\beta=10$ (left) and
 $tan\beta=50$ (right). Black dots denoted models for which $\Omega h^2<.129$.
 The line in the bottom figures is the expected limit from \picasso.
 }} \vspace{-.2cm}
\end{center}
\end{figure*}

We have performed scans over the parameter space of the mSUGRA
model with $\mu>0$, $A_0=0$, $\m0<5$TeV and $\mhf<2$TeV  for
$\tan\beta=10$ and $\tan\beta=50$ as in previous sections. We
first consider a top quark mass $m_t=175$GeV. While the prediction
for the neutralino proton cross sections vary over several orders
of magnitude
 over  the full parameter space, after imposing the WMAP upper
 bound one distinguishes easily  the stau coannihilation region, the light Higgs annihilation region and the focus point region. In the latter case the cross sections are
 typically larger despite the suppressed contribution of the squark exchange.
 The reason is the enhanced coupling of the neutralino to the Higgs
 due to the  mixed Higgsino/gaugino nature of the neutralino. Note  that
the same mechanism that made for efficient neutralino annihilation also induces large values for the spin independent cross section.
 As discussed previously, the focus point region  is found only at large $\tan\beta$ in mSUGRA models  for $\mt=175$~GeV.
 In Fig.~\ref{dd_sugra} the focus point region for $\tan\beta=50$ corresponds to the set of points clustered around $\mneuto=300$~GeV for which
 $\sigma_{SI}\approx 1-3\times10^{-8}$pb.
This region can then be partly probed by \edelweiss II and will be
completely covered by future detectors such as \zepliniv. In
general, one also gets the contribution of the squark exchange
diagram. Thus  large cross sections are also expected in the low
$\m0$ region of parameter space where one finds the  lighter
squarks,
 this basically means the stau coannihilation region.
Furthermore, the spin-independent cross section should
increase for lighter neutralinos as they are found in the low
$\m0-\mhf$ region where  in addition to having light squarks  the coupling of
the neutralino to the  Higgs is large as well. Note however that this is
precisely where  the collider constraints come into play.
After imposing all constraints we find the maximum value
$\sigma_{SI} < 3\times10^{-9}(10^{-8})$pb for $\tan\beta=10(50)$.
These can only be partially probed by future large scale detectors.
For example, \xenon~ will probe models for which $\mneuto <250 (360)$~GeV.

At large $\tan\beta$ there is also a region where neutralinos
annihilate via a heavy Higgs exchange, in this region, neutralinos
are heavy as well as the  squarks and the heavy Higgs which lie in
the TeV range. One therefore does not expect a large cross section
despite the enhanced coupling of the Higgs to bottom quarks.
Indeed this region features cross sections ${\cal{O}}(10^{-10})$pb
and shows up as the continuation of the coannihilation region for
$\mneuto\approx 500$~GeV in Fig.~\ref{dd_sugra}. Finally the
region at intermediate $\m0$ where it is still possible to have
annihilation of neutralinos via the exchange of a light Higgs
scalar predict  values within the range
$10^{-10}(10^{-9})<\sigma_{SI}< 10^{-9}(10^{-8})$~pb for
$\tan\beta=10(50)$. The enhancement at large $\tan\beta$ is due to
the  heavy Higgs contribution. While the heavy Higgs  exchange is
completely negligible at intermediate values of $\tan\beta$ it
eventually dominates the light Higgs despite the mass suppression
factor. This is due partly  to a decrease of the mass of the Heavy
Higgs in models with large $\tan\beta$   but mainly  one benefits
from the enhanced couplings of the heavy Higgs to bottom quarks.
This region is accessible to future large scale detectors such as
\xenon\cite{xenon}.

The spin dependent cross section  has roughly the same behaviour
 as the spin independent one. Indeed the neutralino LSP couplings to the Z is also enhanced in the focus point region because of the larger Higgsino component. This region corresponds to the sest of points clustered around $\sigma_{SD}\approx 10^{-4}$~pb in Fig.\ref{dd_sugra}d and can be probed
 entirely by future large scale detectors such as \picasso\cite{picasso}.
In the coannihilation region and the heavy Higgs annihilation region, the cross section decreases with an increase of the neutralino mass. As above this is related to the additional suppression of the heavy squark exchange.
The predictions  for the spin dependent cross sections still lie orders of magnitude below present and future  limits
especially in the heavy Higgs annihilation region at $\tan\beta=50$ where
$\sigma_{SD} \approx 10^{-8}$~pb. Finally the allowed models with light neutralinos for which a good relic density is found since neutralino annihilates via the light Higgs have also
an enhanced coupling to the Z.  Here one finds rather favourable cross sections, $\sigma_{SI}\approx 10^{-4}-10^{-5}$. These models can be probed by a ton-scale detector.

The neutralino proton cross sections should not depend
very much on the value of the top quark mass for a given set of
MSSM parameters. However the allowed parameter space and the predictions for the MSSM parameters within the mSUGRA model
differs significantly as discussed in section \ref{results}.
If we increase the top quark mass, $\mt=179$~GeV, the focus point region will be shifted to higher values of $\m0$, out of the range scanned here.
Furthermore the heavy Higgs annihilation region almost disappeared. We then recover the results for the spin independent cross section for $\mt=175$~GeV in the coannihilation and  Higgs annihilation regions. However since
the  Higgs mass constraint is  relaxed
one can reach higher  cross sections for light neutralinos, for example  $\sigma_{SI} \approx 5\times10^{-8}$pb for  $\tan\beta=10$.
For  $\tan\beta=50$ the maximum cross section in the coannihilation region
is similar to the case of the lighter top quark since the constraints from colliders are much less dependent on the top quark mass.

\begin{figure*}[tbhp]
\begin{center}
\vspace{-.2cm}
\includegraphics[width=14cm,height=10cm]{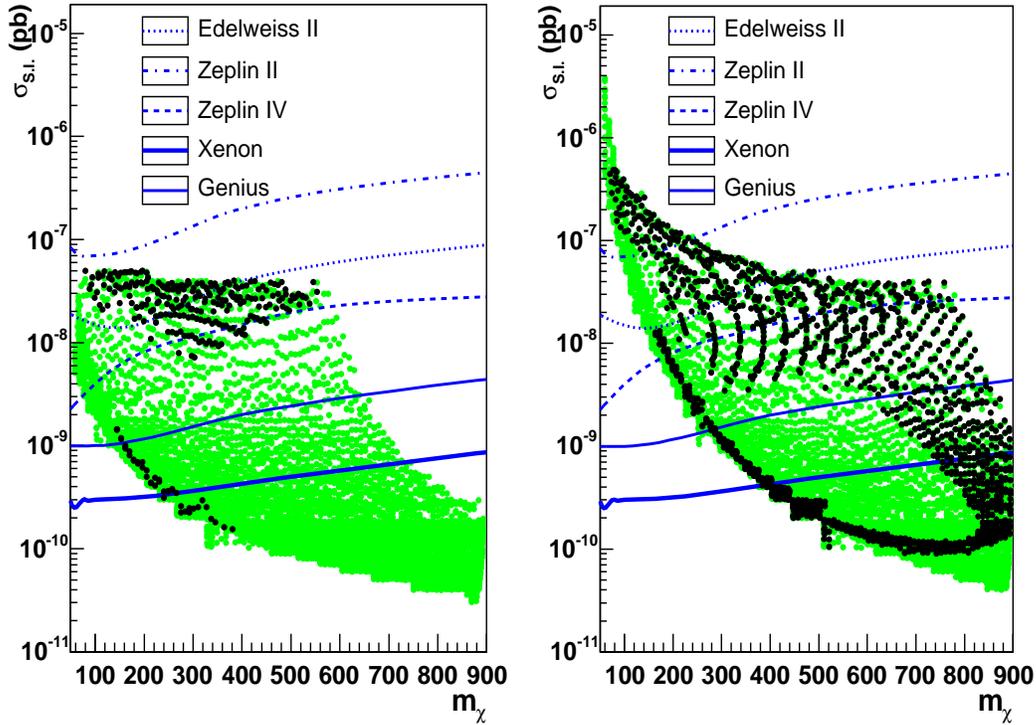}
\vspace{-.1cm} \caption{\label{dd_mt170}{\em Spin independent
cross sections for scattering of neutralinos on proton as fonction
of $\mneuto$ for $\mt=170$~GeV, $\tan\beta=10$ (left) and
 $tan\beta=50$ (right). Black dots denoted models for which $\Omega h^2<.129$
 }} \vspace{-.2cm}
\end{center}
\end{figure*}

We also discuss the case of a rather light top quark since
  WMAP allows a more important region of
parameter space and we can better illustrate the generic behaviour of the neutralino proton cross section in the  different regions
of mSUGRA models.  The results of a scan over models with $m_t=170$GeV
are displayed in  Fig.~\ref{dd_sugra}. The focus point region with a Higgsino LSP  is allowed even for  $\tan\beta=10$, and as above
that is where one finds the largest cross sections,   $\sigma_{SI}\approx
 10^{-8}-10^{-7}$pb  over the allowed range of neutralino masses.
 Much lower cross sections are found in the coannihilation region.
 For  $\tan\beta=50$, the allowed region covers most of the parameter space
 (see Fig.\ref{mt170}b). Nevertheless one can roughly distinguish the focus point, coannihilation and Higgs annihilation regions in Fig.\ref{dd_mt170}b.
  The largest cross sections
are again  found  in the focus point region where  for light
neutralinos $\sigma_{SI}$ can reach  $10^{-7}$pb. The
maximum value for the spin-independent cross section increases
with $\tan\beta$ since one finds a neutralino LSP with a higher
Higgsino content.
In the coannihilation region which shows up as the lower branch in  Fig.~\ref{dd_mt170}b,
 the upper limit drops to $\sigma_{SI}< 10^{-8}$pb
 after imposing the constraints from colliders.
 In the region of the Higgs funnel where neutralinos are above
$500$ GeV,    the spin-independent  cross section drops to rather low values,
 $\sigma_{SI}>7\times 10^{-11}$~pb, beyond the reach of future detectors.
%{\footnote{Note that the heavy Higgs funnel can be recovered even for
%large $\tan\beta$ if one changes the valus of $\mbmb$.}
Basically the conclusions for  the direct detection potential are similar to the ones reached for $\mt=175$~GeV, the best prospects are in the focus point region and for light neutralinos in the coannihilation region while
the heavy neutralinos found in either the Higgs funnel or the coannihilation
region will remain out of reach.
%Note also that the detectors have a reduced sensitivity
%when  the LSP becomes very heavy.

To summarize the results in the  mSUGRA model, the most promising
models for both spin independent and spin dependent direct
detection are models with a mixed bino/Higgsino LSP. For a top
quark mass above 175~GeV, the former is expected only at large
$\tan\beta$. Models with almost a pure bino LSP predict lower
cross sections but should still be within reach of the large scale
detectors provided neutralinos
 do not exceed roughly 400 GeV. Models that belongs to the Higgs funnel region cannot be probed even at the large scale detectors.

\subsection{Non-universal models}

We next discuss non-universal models, considering  the two cases
that differ significantly from mSUGRA as far as the relic density
is concerned: the case with a  LSP with a higher Higgsino content,
$M_3<M_2$ as well as the case with a wino LSP, $M_1>M_2$. Since
the LSP coupling to the Z and/or Higgs is a crucial parameter in
evaluating the neutralino proton cross section one expects
significant differences with the mSUGRA model.
%In non-unified models, one also finds that the  more favourable
%cross sections correspond to  neutralinos around 100GeV where detectors
%have the highest sensitivity.

\subsubsection{$M'_3=M'_2/2$}
\begin{figure*}[tbhp]
\begin{center}
\vspace{-.2cm}
\includegraphics[width=14cm,height=10cm]{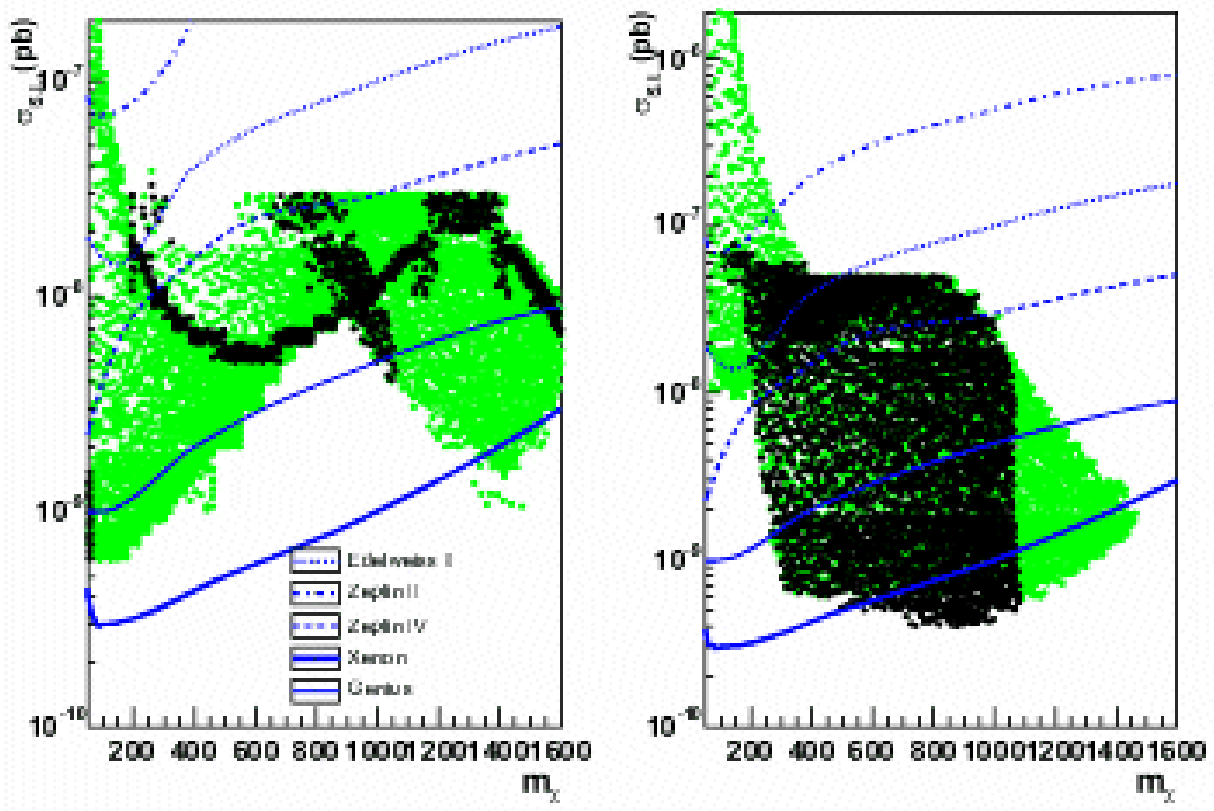}
\vspace{-.2cm} \caption{\label{sim3m22}{\em Spin independent (top)
cross sections for scattering of neutralinos on proton as fonction
of $\mneuto$ for $\tan\beta=10$ and
 $tan\beta=50$ in models where $M'_3=M'_2/2$.
 }} \vspace{-.2cm}
\end{center}
\end{figure*}

First consider the model $r_{32}=2$ and $\mt=175$GeV. We have
performed a scan over the parameters $M'_2<4$TeV and $M_0<5$TeV.
 As we have discussed before,
both the squarks and the heavy Higgs tend to be lighter than in
universal models. Furthermore light neutralinos have a higher
Higgsino component since $\mu$ is typically smaller. Therefore one
expects an enhanced scalar cross section. In Fig.\ref{sim3m22} we
indeed see that in the coannihilation region the lower bound on
the scalar cross section is around $5\times 10^{-9}$~pb and that it
can exceed $\sigma_{SI}=10^{-8}$pb  for $\tan\beta=10$  for light
squarks and neutralinos. The coannihilation/heavy Higgs region
correspond to the inverted  S-shape region in Fig.~\ref{sim3m22}a.
The enhancement of the cross section relative to the universal
model is especially noticeable for heavy neutralinos.
This is solely due to the enhanced coupling of the neutralino to the light Higgs.
 We had found  two focus point regions in  Fig.~\ref{m3m22}.   In Fig.~\ref{sim3m22},   the first region
 corresponds to the sets of points around $\mneuto\approx 200$~GeV
for which $\sigma_{SI}\approx 10^{-8}$, the second region to
$\mneuto\approx 800-1200$~GeV. Towards the upper range of this
region the cross section drops to $\sigma_{SI} \approx 4\times
10^{-9}$. As in the focus point region of the universal model,
reasonable cross sections are found despite the very heavy squarks
because of the mixed bino/Higgsino nature of the LSP. There is
finally one region where one finds  $\sigma_{SI}\approx
10^{-9}$pb, there  neutralinos are light and annihilate through
the light Higgs. We had already found a similar region in mSUGRA
with roughly the same cross section. In this model,  for
$\tan\beta=10$, detectors such as \edelweiss~ probe part of the
focus point region and almost  all regions are within reach of
future detectors like \genius.

For $\tan\beta=50$, most of the parameter space is allowed by the
relic density constraint. The predictions  for $\sigma_{SI}$ spans
a wide range of values. The largest cross sections are found in
the low $\m0-\mhf$ plane, however once the limit on the Higgs mass
and on $\bsgamma$ are taken into account one  finds
$\sigma_{SI}<8\times 10^{-8}$pb. Many models have
$\sigma_{SI}<10^{-9}$~pb, the absolute lower bound is $3\times
10^{-10}$pb, an enhancement as compared to universal models. This is again directly related to the value of $\mu$ which entails a LSP with a higher Higgsino content than in mSUGRA. Some
scenarios can be probed by \edelweiss II but only if
$\mneuto<400$GeV. To probe a large fraction of the
parameter space  one has to wait for large detectors like \xenon,
even then it will not be possible to cover the full parameter
space.

\subsubsection{$M'_1=1.8M'_2$}

When $M'_1=1.8M'_2$ and the LSP is wino-like, the  region where the
WMAP upper bound is satisfied includes all models with
$\mneuto<600$GeV. There the spin independent cross section ranges
from $\sigma_{SI}\approx 10^{-9}- 5\times 10^{-8}$~pb for $\tan\beta=10$.
For a given neutralino mass, the larger values for the cross
sections, as usual, correspond to
 the lighter squarks.
For heavier neutralinos, the WMAP allowed region is one of
gaugino coannnihilation and heavy Higgs annihilation. Although coannihilation helps reduce the
 relic density it does not impact directly on the neutralino proton cross section, one finds small
  cross sections which further
 decrease as one  increases $\mneuto$.
Note that for a fixed value of the neutralino mass in the
coannihilation/Higgs funnel region,  the points that are allowed
by WMAP have the lowest direct detection cross section since they
correspond to larger values of $\m0$. For $\tan\beta=10$,
detectors like \edelweiss II can probe some of the models but even
with \xenon, models with heavy neutralinos cannot be probed. For
$\tan\beta=50$, the prospects for detection by large scale
detectors are better, covering all the
 allowed parameter space. The main reason is that models with heavy neutralinos can only be found in
  the focus point region where the Higgs and Z exchange contribution is important.

\begin{figure*}[tbhp]
\begin{center}
\vspace{-.2cm}
\includegraphics[width=14cm,height=10cm]{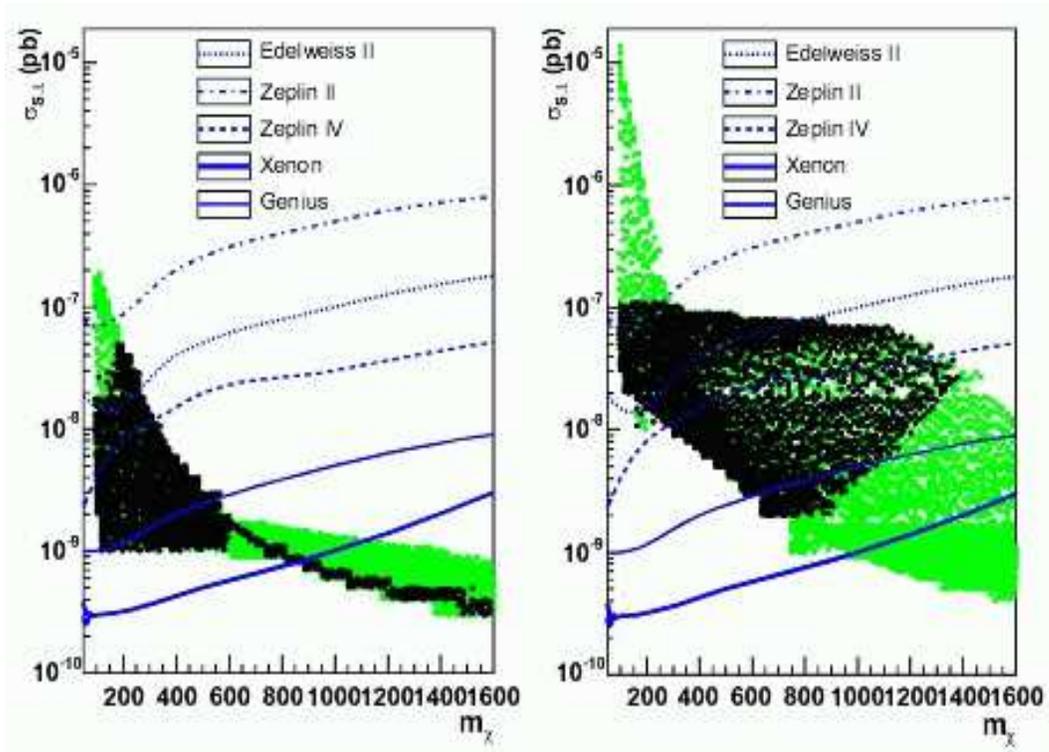}
\vspace{-.2cm} \caption{\label{dd_m118m2}{\em Spin independent
cross sections for scattering of neutralinos on proton as fonction
of $\mneuto$ for $\tan\beta=10$ and
 $tan\beta=50$ in models where $M'_1=1.8M'_2$.
 }} \vspace{-.2cm}
\end{center}
\end{figure*}

\subsection{Summary}

In summary in the focus point region of mSUGRA the prospects are
good for direct detection of dark matter
 by the next generation of  detectors such as \edelweiss II.
In the stau coannihilation region the situation is  more difficult
and one will have to wait for large scale detectors such as
\xenon. Even then only a fraction of the parameter space can be
probed. The Higgs funnel region is extremely dificult even for the
most ambitious project as this region is also characterized  by
heavy neutralinos and rather heavy squarks, one has to face the
problem of reduced sensitivity of detectors even when  the cross
section is not suppressed significantly. In models with
non-universal gaugino masses $M_3<M_2$, prospects are better since
one finds larger values for the neutralino couplings to the Z and
the light Higgs. The lighter squarks also  tend to increase the
cross sections for a given neutralino mass. In models where
$M_1>M_2$ also the prospects for direct detection are more
promising except in  the coannihilation/Higgs funnel region at low
$\tan\beta$. At large $\tan\beta$ the full parameter space can be
basically covered since the  models with heavy neutralinos and
squarks  that are hard to probe  are excluded.

\section{Conclusion}

We have presented constraints on both mSUGRA and non-universal
gaugino masses models emphasizing the role of the latest
measurements of WMAP on the relic density of dark matter. Our
calculations are based on \micro~ coupled to \softsusy~ for the
evaluation of the spectrum.

We found, in agreement with previous analyses, that the new
results of WMAP have almost excluded the bulk region at low
$\m0-\mhf$ in mSUGRA and that it imposes very specific relation
among sparticle masses and parameters:

\begin{itemize}
\item{} Neutralino nearly degenerate with a slepton,
\item{} Neutralino mass near half
the mass of a Higgs,
\item{} Neutralino with a significant Higgsino
component.
\end{itemize}

The latter is  found in the focus point region although the
location of the focus point region depends strongly  on the value
of the top quark mass and to a lesser extent on the choice of
$A_0$. This comes as no surprise since the  solution of the RG
equation for $\mu$ are very sensitive
 to the top Yukawa coupling. Furthermore the different implementations of the higher loop corrections
 in the RGE means that there are still differences
between the predictions of different codes. In our case we found
that a  top quark mass above 179~GeV  pushes the focus point above $\m0=5$~TeV. One
is then basically left with  the coannihilation region as well as
a Higgs annihilation region. The latter includes   a  possibility
for annihilation near a light Higgs resonance but annihilation
through a heavy Higgs resonance is more likely to occur at large
$\tan\beta$. Conversely if the top quark mass lies towards the
lower end of the allowed range,  a large focus point region is
cosmologically allowed even at low
 $\tan\beta$. On the other hand the constraint from the light
 Higgs mass is more severe especially at low values of
 $\tan\beta$.
Note that regions where the stop is the NLSP are usually ruled out
by direct constraints, still the possibility exists in models with
a large mixing in the stop sector.

These conclusions result, in some sense,  from the relation between
the parameters of the MSSM obtained in the mSUGRA model. In order
to assess how general the statements about the importance of the
relic density constraints are, we have moved away from mSUGRA by
introducing some non-universality in the gaugino sector. By doing
so one can deviate
 from mSUGRA predictions for the parameters that are critical in the evaluation
of the relic density in particular  the value of $\mu$.
 This parameter in relation to $M_1,M_2$
 determines the  gaugino/Higgsino content of the LSP and as a result directly the annihilation cross sections into fermions and gauge bosons.
The impact of this parameter is already visible in mSUGRA where as
soon as one could get away from the pure bino case, one finds (in
the focus point region) much more efficient annihilation of
neutralinos. In non-universal MSSM we have, in addition to a
Higgsino LSP, models  with a wino LSP. In these models  the upper
bound on the relic density can easily be satisfied, because of the
large cross section for the annihilation of neutralinos  into W
pairs. These models are the ones where $M'_1>M'_2$ at the GUT
scale. Second, by changing the relation between the heavy Higgs
mass and the neutralino mass we have found that annihilation
through a heavy Higgs exchange can take place even at low
$\tan\beta$. These conditions can be realized in non-universal
gaugino mass models,
 by changing the mass relation between $M'_3$ and $M'_2/M'_1$.
 Indeed $M_A$ is sensitive to the value of $M'_3$ at the high scale.
Finally in all models we found a region where the WMAP upper limit
could be satisfied because of the contribution of slepton
coannihilation channels as long as the neutralino and stau were
not too heavy. In summary in non-universal gaugino mass models,
compatibility with the WMAP measurement  imposes similar
conditions on the supersymmetric masses and parameters than in
mSUGRA. The only new possibility is for a wino LSP in addition to
the  Higgsino LSP. We stress that SUGRA models with non-universal
gaugino masses give a good picture of the dark matter constraints
in the general MSSM while keeping with a reasonable number of free
parameters. Although we have mainly studied only two non-universal
gaugino  mass models, it was sufficient to illustrate the main
mechanism at work in the relic density.

 We have also discussed the potential for probing SUGRA models with
 the different experiments for direct detection of the LSP.
While few models lead to prediction for the direct annihilation cross section within
reach of the current experiments, a large fraction of the parameter space will be probed
by ton-scale future detectors.
 Typically models with a Higgsino LSP predict the largest cross sections
 while coannihilation models feature much lower cross sections,
 especially when the sfermion masses
 increase. In fact the Higgsino component of the LSP is also what makes efficient annihilation of
 neutralino pairs and leads to acceptable values for the relic density. Finally when the neutralino annihilates via a
  heavy Higgs, few models can lead to a signal for direct detection  especially when they  feature a  heavy LSP.
 We keep a more detailed comparative study of the direct detection and collider searches for a future
 publication.

\section{Acknowledgements}

We thank B. Allanach for discussions and help with the \softsusy~
code.  This work was supported in part by the PICS-397 of CNRS, {\it Calcul en physique des particules}, by GDRI-ACPP of CNRS and by grants from the
Russian Federal Agency for Science, NS-1685.2003.2  and RFBR 04-02-17448.

%\bibliography{bankspires}

\providecommand{\href}[2]{#2}\begingroup\raggedright\endgroup

\end{document}